\documentclass[11pt,a4paper]{article}

\usepackage[utf8]{inputenc} 
\usepackage[T1]{fontenc}    %
\usepackage[english]{babel} 
\usepackage[final]{microtype} 

\usepackage{amssymb,amsmath,mathtools} 
\usepackage{amsthm} 

\usepackage{xcolor}
\usepackage{bbm}
\usepackage{braket}

\usepackage[hmargin=0.12\paperwidth,vmargin=0.19\paperwidth,bindingoffset=0cm]%
{geometry} 

\numberwithin{equation}{section} 

\theoremstyle{plain}
  
  \newtheorem{prop}{Proposition}
  
  \newtheorem{cor}{Corollary}
\theoremstyle{definition}
  
  \newtheorem{remark}{Remark}
  
  \numberwithin{prop}{section}
   \numberwithin{cor}{section}
   \numberwithin{remark}{section}

\title{\Large\bfseries Computing marginal eigenvalue distributions \\%
    for the Gaussian and Laguerre orthogonal ensembles}
\author{Peter J. Forrester${}^{(1)}$, Santosh Kumar${}^{(2)}$
and Bo-Jian Shen${}^{(1)}$}
\date{}

\begin{document}

\maketitle

${}^{(1)}$
School of Mathematics and Statistics,  The University of Melbourne,
Victoria 3010, Australia. \: \: Email: {\tt pjforr@unimelb.edu.au}; \,
{\tt bojian.shen@unimelb.edu.au}\\

${}^{(2)}$\footnote{Deceased 18/10/24}
Department of Physics, Shiv Nadar Institution of Eminence, Gautam Buddha, Uttar Pradesh - 201314,
India.\: \: Email: {\tt skumar.physics@gmail.com}

\bigskip

\begin{abstract}
\noindent 
 The Gaussian and Laguerre orthogonal ensembles are fundamental to random matrix theory, and
the marginal eigenvalue distributions 
 are basic observable quantities. Notwithstanding a long history, a formulation providing high precision numerical
 evaluations for $N$ large enough to probe asymptotic regimes, has not been provided.
 An exception is for the largest eigenvalue, where there is a formalism due to Chiani which uses a combination of the
 Pfaffian structure underlying the ensembles, and a recursive computation of the matrix elements. We augment this
 strategy by introducing a generating function for the conditioned gap probabilities. A  finite Fourier series approach is then used to extract the sequence of marginal eigenvalue distributions as a linear combination of Pfaffians, with the latter then evaluated using
 an efficient
 numerical procedure available in the literature. Applications are given to illustrating 
 various asymptotic formulas, local central limit theorems, and central limit theorems, as well as to probing finite size
 corrections. Further, our data indicates that the mean values of the marginal distributions interlace with the zeros of the Hermite polynomial (Gaussian ensemble) and a Laguerre polynomial (Laguerre ensemble).
\end{abstract}

\vspace{3em}

\section{Introduction}
\subsection{Background}

Let $G_{n,N}$ denote an $n \times N$ rectangular standard real Gaussian matrix, meaning that each entry is chosen
independently as a standard normal random variable. With $A_N = G_{N,N}$ and $B_{n,N} = G_{n,N}$ $(n \ge N)$ being
particular square and rectangular real Gaussian matrices, introduce the real symmetric matrices
\begin{equation}\label{a1.0}
X_N = {1 \over 2} (A_N + A_N^T), \qquad Y_{n,N} = B_{n,N}^T B_{n,N}.
\end{equation}
Random matrices with the construction of $X_N$ are said to form the Gaussian orthogonal ensemble (GOE) \cite[\S 1.1]{Fo10}.
 Random matrices with the construction of $Y_N$ are 
 examples of uncorrelated real Wishart matrices from multivariate statistics \cite[\S 3.2]{Fo10},
 and through their eigenvalue PDF relate to the
  Laguerre orthogonal ensemble (LOE); see below.
 
 Standard results in random matrix theory \cite[Prop.~1.3.4 and 3.2.2]{Fo10} give that the eigenvalue probability density functions (PDFs)
 for the random matrices (\ref{a1.0}) are
 \begin{align}
 \mathcal P^{\rm G}(x_1,\dots,x_N) & = {1 \over Z_N^{\rm G}} \prod_{l=1}^N e^{- x_l^2/2} \prod_{1 \le j < k \le N} | x_k - x_j| \label{a1.1}
 \\
  \mathcal P^{\rm L}(x_1,\dots,x_N) & = {1 \over Z_N^{\rm L}} \prod_{l=1}^N x_l^a e^{- x_l/2} \mathbbm{1}_{x_l > 0}  \prod_{1 \le j < k \le N} | x_k - x_j|, \quad
  a=(n-N-1)/2 \label{a1.2},
  \end{align}
  with normalisations
  \begin{equation}
  Z_N^{\rm G} = \prod_{j=1}^N \Gamma(1 + j/2), \qquad
  Z_{n,N}^{\rm L} = N! \pi^{-N/2} 2^{Nn/2} \prod_{j=1}^{N} \Gamma(j/2) \Gamma ((j + n - N )/2).
  \end{equation}
  In the above, we have used the subscripts G (L) in relation to the GOE (LOE). One should mention at this stage that the
  terminology LOE applies more generally when the exponent
   $a=(n-N-1)/2$ in each factor $x_l^a$ of (\ref{a1.2}) is regarded as a continuous parameter $a > -1$. Random matrices that
   realise this more general eigenvalue PDF are known \cite{DE02}, but they cannot be constructed out of matrices with standard 
   Gaussian entries.

  Our interest in this paper is in the computation  of the individual eigenvalue marginals corresponding to (\ref{a1.1}) and
  (\ref{a1.2}), and the underlying conditioned gap probabilities. The eigenvalue marginals are the PDFs $f_N^{(\cdot)}(k;s)$ for the $k$-th largest eigenvalue $(k=1,\dots,N)$.
  As is well known \cite[\S 8.1]{Fo10}, these PDFs can be expressed in terms of $\{E_N^{(\cdot)}(j;(s,\infty)) \}_{j=0}^{k-1}$, 
  where $E_N^{(\cdot)}(j;(s,\infty))$ denotes the conditioned gap probability that there are exactly $j$ eigenvalues in the interval $(s,\infty)$, and
  which themselves
  specify the distribution  of the random variable $\sum_{l=1}^N \mathbbm 1_{x_l > s}$
  corresponding to the number of eigenvalues in the half interval $(s,\infty)$.
   Specifically, one has the recursive formula
   \begin{equation} \label{a1.3}
   f_N^{(\cdot)}(k;s) = {d \over d s} E_N^{(\cdot)}(k-1;(s,\infty)) + f_N^{(\cdot)}(k-1;s) 
   \end{equation}
   subject to the initial condition $ f_N^{(\cdot)}(0;s) = 0$, telling us that
    \begin{equation} \label{a1.3i}
    f_N^{(\cdot)}(k;s) = {d \over d s} \sum_{l=0}^{k-1}  E_N^{(\cdot)}(l;(s,\infty)).
      \end{equation}
      
   Regarding motivation, one notes that 
   the individual eigenvalue marginals for the GOE appear explicitly in
   formulas relating to critical points of isotropic Gaussian random fields \cite{ABC13,AD22}. Further, associated with individual
   eigenvalues, and for the conditioned gap probabilities, are various limit laws. While for the former these are best known at the spectrum edge \cite{Fo93a,TW94a}, there are also limit
   laws of individual eigenvalues in the bulk, and moving inwards from the edge with the matrix size \cite{Gu05,OR10}. This is similarly true of the conditioned
   gap probabilities, or more specifically the random variable for the number of eigenvalues in an interval, in the setting that the number of eigenvalues
   in the interval tends to infinity \cite{CL95,So00,Ki08}.
   
   Results in the literature have addressed the computation of (\ref{a1.3}) for the eigenvalue PDFs (\ref{a1.1}) and (\ref{a1.2}) in
   two distinct, fundamental ways. The first line of study aimed to give exact functional forms for small values of $N$. The tool for this was the discovery
   by Davis \cite{Da72a,Da72b} that associated with $\{ f_N^{(\cdot)}(k;s) \}$ is an $(N+1) \times (N+1)$ first order matrix differential equation,
   which allows for a recursive computational scheme.
   In  \cite[Appendix B]{Da72b} there is a listing of exact functional forms of $\{ f_N^{\rm L}(k;s) \}_{k=1}^N$ up to $N=5$. These ideas were
   developed in the Gaussian case (\ref{a1.1}) in the PhD thesis of Eckert \cite{Ek74} (this reference is available in electronic form on the
   internet),
   which contains the evaluations of   $\{ f_N^{\rm G}(k;s) \}_{k=1}^N$ up to $N=7$.   
   
   Another  of the results of
    \cite{Ek74}, was the identification of a transcendental basis for the PDF $ f_N^{\rm G}(0;s)$   (a summary is given in \cite{Ja75}). To specify this basis,
    write 
     $\phi(s) = e^{-s^2/2}$, $\Phi_1(s) = \int_{-\infty}^s \phi(y) \, dy$,  $\Phi_2(s) = \int_{-\infty}^s (\phi(y))^2 \, dy$, and define functions 
     $\{ \Omega_j(s) \}$ 
    \begin{equation} \label{tb1}   
    \Omega_{2k+1}(s) = \Phi_1(s)  (\Phi_2(s))^k \: \: (k=0,1,\dots), \qquad   \Omega_{2k}(s) =  (\Phi_2(s))^k  \: \: (k=0,1,\dots).
    \end{equation}
    With this notation, it is established in   \cite[after the change of variables $s \mapsto \sqrt{2} s$]{Ek74}  that
     \begin{equation} \label{tb2}   
      f_N^{\rm G}(1;s) =  \sum_{j=0}^{N-1}  \pi_j(s)  ( \phi(s) )^{N-1 -j} \Omega_j(s)
     \end{equation}
     for some polynomials $\{ \pi_j(s) \}$. Moreover, a family of linear operators $\{ T_p \}_{p=1}^{N-1}$ was identified with
     an explicit action on any functional form  with the structure of the right hand side of (\ref{tb2}), and
     which allows for each member of $ \{ f_N^{\rm G}(n;s) \}_{n=2}^N$ to be deduced using  knowledge of (\ref{tb2}).
     For the Laguerre case (\ref{a1.2}) it was shown in \cite{FK19} 
     for $k=1$, and subsequently in \cite{FK24} for general $k$ ($k=1,\dots,N$),
     that with
     $a:=(n-N-1)/2$ a non-negative integer, there are polynomials $q_j(s;k,N,a)$ of degree $j(a+N-j)$ such that
     \begin{equation} \label{tb3}   
        f_N^{\rm L}(k;s) =   \sum_{j=n}^{N}  q_j(s;k,N,a) e^{-j s/2}; 
     \end{equation}
     see too \cite{MP24}.
     In addition, the references   \cite{FK19}  and  \cite{FK24} implement in Mathematica notebooks the recursive method of Davis
     which provides for the exact evaluation of the polynomials. Regarding efficiency, in the case $k=1$  and $a=0$, $N=40$ the
     evaluation of $\{  q_j(s;k,N,a) \}_{j=1}^N$ takes  around 21 minutes using a 2017 model IMac machine. 
     However the recurrence is such that to compute
     $\{ f_N^{\rm L}(k;s) \}_{k=1}^m$ requires having first computed  $\{ f_M^{\rm L}(k;s) \}_{k=1}^{{\rm max} \{m,M\}}$ for $M=1,\dots,N-1$.
     While the algorithm is only cubic in $N$, with three main loops, there are overheads associated with the intermediate computer algebra
    specific operations.
     This limits the practical use of the code with respect to the size of $N$ and $M$; for example with $a=0$ and
     $N=M=11$ the run-time is around 93 minutes using the same machine.
     
     A special functional form has also been identified
     in the case that $a+1/2$ is a non-negative integer   \cite{FK24}. This is most conveniently written in terms of
    the cumulative distribution $F_N^{\rm L}(1;x) = \int_0^x     f_N^{\rm L}(1;s)   \, ds$, and depends too on the
    parity of $N$. Setting  $N$ even for convenience, it was exhibited in  \cite{FK24} 
    for small  values of $N$ and conjectured
to hold in general that there are polynomials $p_{l,1}^{\rm e}(x), p_{l,2}^{\rm e}(x)$ such that
   \begin{equation} \label{tb4}      
F_N^{\rm L}(1;x) =
\sum_{l=1}^{N/2+1} \Big (  e^{-(l-1)x} p_{l,1}^{\rm e}(x) +
 \sqrt{x} {\rm erf}(\sqrt{x/2}) e^{-(l-1/2)x}
p_{l,2}^{\rm e}(x) \Big ).
 \end{equation}

The second fundamental approach to the computation of $   f_N^{(\cdot)}(k;s)$  available in the literature --- albeit
presented only in the case $k=1$ --- is due to Chiani \cite{Ch14}. The starting point for this is the classical
knowledge in random matrix theory due to Mehta \cite{Me67}, that computing averages with respect to (\ref{a1.1})
gives rise to a Pfaffian structure, and which moreover remains true in the Laguerre case (\ref{a1.2}).
This is a consequence of the method of integration over alternate variables, which can be traced
back to de Bruijn \cite{dB55}. For the average corresponding to $E_N^{(\cdot)}(0;(s,\infty)) $ in both the Gaussian and Laguerre
cases it was shown in  \cite{Ch14}  that the matrix elements of the Pfaffian, which at first are given as double integrals, can
be calculated recursively starting from a particular special function (error function in the Gaussian case, incomplete
gamma function in the Laguerre case). This allows for an efficient computation of the matrix elements to high precision,
with the later feature carrying over to the numerical evaluation of the Pfaffian, and leading via an application of (\ref{a1.3}) in the
case $n=1$ to a high precision evaluation of $   f_N^{(\cdot)}(1;s)$. The advantage of this approach relative to the one based on
exact functional forms is that larger values of $N$ can be accessed. For example, in Chiani \cite{Ch14} a run-time of just 5
seconds is reported for the Laguerre case with $a=0$, $N=200$, albeit for a fixed value of $s$.

\subsection{Specific aims and paper outline}
As already stated, our primary concern in this paper is with  the computation  of the individual eigenvalue marginals for the
joint PDFs (\ref{a1.1}) and
  (\ref{a1.2}), and the underlying conditioned gap probabilities. 
  We will proceed using the (well known --- recall the final paragraph above) Pfaffian structure, combined with
  the insight from \cite{Ch14} that the matrix elements can be computed by recurrence. The computer algebra system Mathematica
  allows for a high precision computation of these matrix elements. Essential too is use of the
  Mathematica code of Wimmer \cite{Wi12} for the efficient numerical evaluation of a Pfaffian.
  
  The details of the recurrences for the Gaussian and Laguerre cases differ, necessitating that they be treated
  separately. Also, the feature of the GOE of being independent of a parameter beyond the matrix size
  gives rise to unique marginal probability density functions. For these, following on from work began in
  the  thesis of Eckert \cite{Ek74}, there is interest in providing high precision statistical data relating to the cumulants.
  Eckert provides such a tabulation up to and including $N=7$. Using our exact (internally stored within Mathematica)
  functional forms obtained from the Pfaffian formulation allows this tabulation to be extended up to and including
  $N=12$ (Appendix A). Unlike the LOE, the GOE is symmetrical about the origin, which distinguishes the conditioned
  probabilities $\{  E_N^{(\rm G)}(k;(0,\infty))  \}_{k=0}^\infty$, which determine the number of positive eigenvalues.
  References relating to the interest in this quantity are given
  in the first paragraph of \S \ref{S3.1}. This provides motivation to investigate  the large $N$ form of the variance of the number
  of positive eigenvalues, and also the accuracy of a known local central limit theorem. In the literature the case $k=0$
  of these conditioned
  probabilities. In particular, there is a known large $N$ asymptotic formula which we are able to illustrate. As two extra applications
  of our numerics in the Gaussian case, we consider the rate of convergence to a known central limit law for the marginal probability density function
  near the centre of the spectrum, and also interlacing properties of the mean of the marginal   distributions (for the latter see Appendix A). Analogous asymptotic questions are probed in the Laguerre case in Section \ref{S3.2}, with an interlacing property of the means of the marginal densities considered in Appendix A.
  
  \section{Pfaffian formulation and exact evaluations}
  Fundamental to our working is the  generating function 
  \begin{equation}\label{1.0}
  \Xi_N^{(\cdot)}((s,\infty);\zeta) := \sum_{k=0}^N \zeta^k E_N^{(\cdot)}(k;(s,\infty)) 
   = \bigg \langle \prod_{l=1}^N \Big (  \mathbbm{1}_{x_l < s }  + \zeta   \mathbbm{1}_{x_l > s } \Big ) \bigg \rangle^{(\cdot)},
  \end{equation}
  where the average is with respect to one of the joint PDFs (\ref{a1.1}) or
  (\ref{a1.2}). A Pfaffian form of the average in (\ref{1.0}) can be written down for any joint PDF structurally identical
  to (\ref{a1.1}) and
  (\ref{a1.2}), although the details depend on the parity of $N$ \cite[Prop.~6.3.4 and Exercises 6.3 q.1]{Fo10}. To specify the Pfaffian,
  introduce
  \begin{align}\label{1.2}  
   & H_0 ^{\rm G}(j,k;s) = {1 \over \Gamma(j/2) \Gamma(k/2)} \int_{-\infty}^s dx \, x^{j-1} e^{-x^2/2}  \int_{-\infty}^s dy \, y^{k-1} e^{-y^2/2} {\rm sgn} (y-x)  \nonumber  \\
    & H_1^{\rm G}(j,k;s) =  {1 \over \Gamma(j/2) \Gamma(k/2)} \int_s^{\infty} dx \, x^{j-1} e^{-x^2/2}  \int_{-\infty}^s  dy \, y^{k-1} e^{-y^2/2} - (j \leftrightarrow  k)  \nonumber \\
    & H_2^{\rm G}(j,k;s) =  {1 \over \Gamma(j/2) \Gamma(k/2)} \int_s^{\infty} dx \, x^{j-1} e^{-x^2/2}  \int_s^{\infty}  dy \, y^{k-1} e^{-y^2/2}    {\rm sgn} (y-x) 
    \end{align}  
    and
 \begin{align}\label{1.3}  
   & H_0^{\rm L}(j,k;s) = {1 \over \Gamma(a+j) \Gamma(a+k)} \int_{0}^s dx \, x^{a+j-1} e^{-x}  \int_{0}^s dy \, y^{a+k-1} e^{-y} {\rm sgn} (y-x)  \nonumber  \\
    & H_1^{\rm L}(j,k;s) =  {1 \over \Gamma(a+j) \Gamma(a+k)} \int_s^{\infty} dx \, x^{a+j-1} e^{-x}  \int_{0}^s  dy \, y^{a+k-1} e^{-y} - (j \leftrightarrow  k)  \nonumber \\
    & H_2^{\rm L}(j,k;s) =  {1 \over \Gamma(a+j) \Gamma(a+k)} \int_s^{\infty} dx \, x^{a+j-1} e^{-x}  \int_s^{\infty}  dy \, y^{a+k-1} e^{-y}    {\rm sgn} (y-x).    
   \end{align} 
   Note that in our notation $H_\mu^{\rm L}(j,k;s)$ the dependence on the parameter $a$ has been suppressed.     Further introduce the column vector
   $[\nu_{j}^{(\cdot)}(s) ]_{j=1}^{N+1}$, where  $\nu_{N+1}^{(\cdot)}(s) = 0$ and
     \begin{align}
    \nu_j^{\rm G}(s) & = {1 \over  \Gamma(j/2) } \bigg (   \int_{-\infty}^s dx \, x^{j-1} e^{-x^2/2} + \zeta  \int_s^{\infty} dx \, x^{j-1} e^{-x^2/2}  \bigg ) \label{1.4a}\\
        \nu_j^{\rm L}(s) & = {1 \over  \Gamma(a+j) } \bigg (   \int_{0}^s dx \, x^{a+j-1} e^{-x} + \zeta  \int_s^{\infty} dx \, x^{a+j-1} e^{-x}  \bigg ),  \label{1.4b}
    \end{align}
    for $j=1,\dots,N$, where here the dependence on $\zeta$ has been suppressed in the notation for both quantities, as well as the dependence on $a$ in $ \nu_j^{\rm L}$.

\begin{prop}\label{P1}
For $N$ even
   \begin{align}\label{1.1} 
    \Xi_N^{\rm G}((s,\infty);\zeta)  & = 2^{N/2} {\rm Pf}  \Big [ H_0^{\rm G}(j,k;s)  + \zeta H_1^{\rm G}(j,k;s) + \zeta^2  H_2^{\rm G}(j,k;s) \Big ]_{j,k=1}^N, \nonumber \\
   \Xi_N^{\rm L}((s,\infty);\zeta)  & =    \prod_{j=1}^N
    {\pi^{1/2} \Gamma(a+j) \over  \Gamma(j/2) \Gamma (a+ (j +1)/2 )}  \nonumber \\
  & \qquad \times  {\rm Pf}   \Big [ H_0^{\rm L}(j,k;s/2)  + \zeta H_1^{\rm L}(j,k;s/2) + \zeta^2  H_2^{\rm L}(j,k;s/2) \Big ]_{j,k=1}^N.
     \end{align}
 In the case of $N$ odd, the size of the matrices that its Pfaffian is being computed in (\ref{1.1}) 
   must be increased from $N$ to $N+1$ by the bordering of the existing matrix by an additional column with entries
   $[\nu_j^{\rm G}(s)]_{j=1}^{N+1}$ and $[\nu_j^{\rm L}(s/2)]_{j=1}^{N+1}$ in the Gaussian and Laguerre cases respectively.
   \end{prop}     

\begin{remark}
For both   (\ref{a1.1}) and
  (\ref{a1.2}) there are what may referred to as Fredholm Pfaffian formulas for $ \Xi_N^{(\cdot)}((s,\infty);\zeta)$
  \cite[Eq.~(180)]{Ma11}, which relate to some specific $2 \times 2$ anti-symmetric integral kernel; see also
  \cite[\S 2.2.2 and \S 2.2.3]{Bo10}. In the circumstance that the integral kernel is an analytic function,
  such Fredholm expressions offer powerful computational properties \cite{Bo08,Bo10}. However, as noted
  in \cite[\S 2.2.3]{Bo10}, for the matrix integral kernel that results from the Pfaffian point processes with orthogonal
  symmetry such as (\ref{a1.1}) and
  (\ref{a1.2}) one of the matrix entries contains the non-analytic additive factor ${\rm sgn}(x-y)$, which nullifies its computational
  value.
  \end{remark}

   The computational scheme for the matrix elements depends on the details of the weight function 
 and so  differs in the Gaussian and Laguerre cases. Therefore each will
   be considered separately.
  
  \subsection{Gaussian case}\label{S2.1}
  Define
  \begin{align}\label{1.4c}
  & \Psi^{\rm G}(j;x)  := {1 \over \Gamma(j/2)} \int_{-\infty}^x t^{j-1} e^{- t^2/2} \, dt, \quad I^{\rm G}(j,k;x) :=  {1 \over \Gamma(j/2)} \int_{-\infty}^x t^{j-1} e^{- t^2/2}   \Psi^{\rm G}(k;t) \, dt \nonumber \\
   & \tilde{\Psi}^{\rm G}(j;x)  := {1 \over \Gamma(j/2)} \int_x^{\infty} t^{j-1} e^{- t^2/2} \, dt, \quad \tilde{I}^{\rm G}(j,k;x) :=  {1 \over \Gamma(j/2)} \int_x^{\infty} t^{j-1} e^{- t^2/2}  \tilde{\Psi}^{\rm G}(k;t) \, dt.
  \end{align}
  Using this notation, the quantities in the Gaussian cases of (\ref{1.2}) and (\ref{1.4a}) can be written
  \begin{align}\label{1.2a}  
   & H_0 ^{\rm G}(j,k;s) = -   I^{\rm G}(j,k;s)  +   I^{\rm G}(k,j;s), \quad
     H_1^{\rm G}(j,k;s) =    -\tilde{\Psi}^{\rm G}(j;s)    {\Psi}^{\rm G}(k;s)  +    \tilde{\Psi}^{\rm G}(k;s)    {\Psi}^{\rm G}(j;s)    \nonumber \\
    & H_2^{\rm G}(j,k;s) =     \tilde{I}^{\rm G}(j,k;s)  -   \tilde{I}^{\rm G}(k,j;s) , \quad    \nu_j^{\rm G}(s)  =  \Psi^{\rm G}(j;s) + \zeta  \tilde{\Psi}^{\rm G}(j;s).
    \end{align}  
    By writing $\tilde{\Psi}^{\rm G}$ in terms of ${\Psi}^{\rm G}$, the expression for $ H_1^{\rm G}$ can be further simplified to read
    \begin{equation}\label{1.2b}  
     H_1^{\rm G}(j,k;s) = -u_j     {\Psi}^{\rm G}(k;s) + u_k  {\Psi}^{\rm G}(j;s),
   \end{equation} 
   where $u_l=0$ for $l$ even and $u_l = 2^{l/2}$ for $l$ odd.    
    Following Chiani \cite{Ch14}, using straightforward integration by parts,  each of the functions in (\ref{1.4c}) can be determined by second order recurrences and appropriate initial conditions, thus
    allowing for the determination of the $H_i^{\rm G}$.
    
    \begin{prop}
    We have the recurrences
     \begin{align}\label{1.3x}  
     & \Psi^{\rm G}(j;x)  = - {x^{j-2} \over \Gamma(j/2)} e^{- x^2/2} + 2 \Psi^{\rm G}(j-2;x), \quad  \tilde{\Psi}^{\rm G}(j;x)  = {x^{j-2} \over \Gamma(j/2)} e^{- x^2/2} + 2 \tilde{\Psi}^{\rm G}(j-2;x),    \nonumber \\
     &  I^{\rm G}(j+2,k;x) = 2  I^{\rm G}(j,k;x) -  {x^{j}   e^{- x^2/2}  \over \Gamma(j/2+1)}  \Psi^{\rm G}(k;x)  + {2^{-(j+k)/2}  \Gamma((j+k)/2) \over  \Gamma(j/2+1)  \Gamma(k/2+1) }
       \Psi^{\rm G}(j+k;\sqrt{2} x),    \nonumber \\
  &  \tilde{I}^{\rm G}(j+2,k;x) = 2  \tilde{I}^{\rm G}(j,k;x) +  {x^{j}   e^{- x^2/2}  \over \Gamma(j/2+1)}  \tilde{\Psi}^{\rm G}(k;x)  - {2^{-(j+k)/2}  \Gamma((j+k)/2) \over  \Gamma(j/2+1)  \Gamma(k/2+1) }
       \tilde{\Psi}^{\rm G}(j+k;\sqrt{2} x),     
     \end{align} 
     where in the first line $ j \ge 2$, while in the second and third line $ j \ge 0$.  Associated  initial conditions are
      \begin{align}\label{1.3a} 
    & \Psi^{\rm G}(0;x)  =     0,      \quad   \Psi^{\rm G}(1;x)  =    {1 \over \sqrt{2}} \Big ( 1 + {\rm erf} (x/\sqrt{2}) \Big ), \nonumber \\
      &\tilde{\Psi}^{\rm G}(0;x)  =     0,      \quad    \tilde{\Psi}^{\rm G}(1;x)  =    {1 \over \sqrt{2}} \Big ( 1 - {\rm erf} (x/\sqrt{2}) \Big ),  \nonumber \\
     &    I^{\rm G}(0,k;x) =   I^{\rm G}(j,0;x) = 0, \quad     \tilde{I}^{\rm G}(0,k;x) =   \tilde{I}^{\rm G}(j,0;x) = 0,  \, j,k \ge 1, \nonumber \\
   &    I^{\rm G}(1,1;x)    = {1 \over 2} \Big ( \psi^{\rm G}(1;x) \Big )^2, \quad
     \tilde{I}^{\rm G}(1,1;x)    = {1 \over 2} \Big ( \tilde{\psi}^{\rm G}(1;x) \Big )^2.
      \end{align}      
     \end{prop}
     
     For a given (even) value of $N$,
      the recurrences in the first line of (\ref{1.3x}) are to be iterated for $j=2,3,\dots,2N$ using the initial conditions in the first  line of (\ref{1.3a}).
      Thus we have available the values of 
      \begin{equation}\label{Pv}  
      \{ \Psi^{\rm G}(j;x) , \tilde{\Psi}^{\rm G}(j;x) \}_{j=0}^{2N}, 
        \end{equation} 
      which are required to implement the
      recurrences for $I^{\rm G}, \tilde{I}^{\rm G}$. In relation to  the latter, as a start
      the initial conditions in the third line of (\ref{1.3a}) with $k=1$ and in the final line, are to be used  in the recurrences of the second and third line of (\ref{1.3x}),
      allowing for the computation of 
     \begin{equation}\label{2.5a}    
     \{ I^{\rm G}(j,1;x),  \tilde{I}^{\rm G}(j,1;x) \}_{j=2}^N.
   \end{equation}
   The quantities $ I^{\rm G}$ and     $\tilde{I}^{\rm G}$ also satisfy the symmetry relations
      \begin{equation}\label{2.5b} 
    I^{\rm G}(j,k;x)   = \psi^{\rm G}(j;x)  \psi^{\rm G}(k;x) -   I^{\rm G}(k,j;x), \quad     \tilde{I}^{\rm G}(j,k;x)   = \tilde{\psi}^{\rm G}(j;x)  \tilde{\psi}^{\rm G}(k;x) -   \tilde{I}^{\rm G}(k,j;x),
      \end{equation}
      the derivation of which is an appropriate integration by parts (in the case of $  I^{\rm G}(j,k;x)$ for example, the starting point is to note that
      $I^{\rm G}(j,k;x) = \int_{-\infty}^x \Big ( {d \over dx}  {\psi}^{\rm G}(j;x) \Big )   {\psi}^{\rm G}(k;x) \, dx$).
      Setting $j=1$ and using knowledge of the values of (\ref{2.5a}) provides the values of
     \begin{equation}\label{2.5c}  
     \{  I^{\rm G}(1,k;x),  \tilde{I}^{\rm G}(1,k;x) \}_{k=2}^N.  
    \end{equation}
    This, together with the $k$-dependent initial conditions in the third line of (\ref{1.3a}), provide initial conditions to use    
   the recurrences of the second and third line of (\ref{1.3x}) to compute
      \begin{equation}\label{2.8}
      \{   I^{\rm G}(j,k;x),  \tilde{I}^{\rm G}(j,k;x) \}_{j=2}^N 
   \end{equation}  
   for each $k=2,\dots,N$. Combining then the evaluations in the final line of (\ref{1.3a}) with  (\ref{2.5c}) and (\ref{2.8}) we have
   the evaluation of the full $N \times N$ array
       \begin{equation}\label{2.9}
      \{   I^{\rm G}(j,k;x),  \tilde{I}^{\rm G}(j,k;x) \}_{j,k=1}^N. 
   \end{equation}  
     
  We see from (\ref{1.2a}) that knowledge of the values of (\ref{Pv}) and (\ref{2.8}) makes explicit all the matrix elements in the
  Pfaffian formula of Proposition \ref{P1} for $ \Xi_N^{\rm G}((s,\infty);\zeta) $. It is well known that for $A$ a $2n \times 2n$
  anti-symmetric matrix $[a_{j,k}]_{j,k=1}^{2n}$, one has for the corresponding Pfaffian the Laplace type expansion
   \begin{equation}\label{2.10} 
  {\rm Pf}( A) = \sum_{r=1}^{2n-1} (-1)^{r+1} a_{r,2n} {\rm Pf}_{r,2n}(A),
     \end{equation}
     where $ {\rm Pf}_{r,2n}$ is the Pfaffian of the $(2n-2) \times (2n-2)$  anti-symmetric matrix  obtained by deleting rows
     and columns $r,2n$; see e.g.~\cite[Exercises 6.1]{Fo10}. Iterative use of this allows for the $(2n-1)!!$ terms
     (each a degree $n$ monomial in the matrix elements $\{ a_{j,k} \}$) in the fully expanded form
     of $  {\rm Pf}( A)$ to be
     made explicit (a brief code in the computer algebra system Mathematica for this purpose is given in \cite{SE16}).
      From (\ref{1.0}), in relation to the first Pfaffian in (\ref{1.1}) this is a polynomial in $\zeta$, the coefficients of which are
     the probabilities $\{E_N^{\rm G}(k;s)\}_{k=1}^N$. With knowledge of the latter, application of the formula (\ref{a1.3}) gives
     $\{f_N^{\rm G}(k;s)\}_{k=1}^N$. The symmetry of the GOE spectrum upon reflection about the origin implies
   \begin{equation}\label{2.11} 
  f_N^{\rm G}(k;s) = f_N^{\rm G}(N+1-k;-s),
  \end{equation}       
and so it suffices to extract only       $\{E_N^{\rm G}(k;s)\}_{k=1}^{\lceil N/2 \rceil}$.

Code has been written in the computer algebra system Mathematica to carry out the above steps. Most time consuming is the
need to use the command {\tt Simplify} on the coefficients of $\zeta$ obtained by the expansion of the Pfaffian. Without
applying this command, as $N$ increases it is not possible to make use of the output for purposes of characterising the statistical properties of
the $f_N^{\rm G}(k;s)$, due to the subsequent failure of the command {\tt NIntegrate}. Such overheads have limited our exact determination
(albeit stored electronically)\footnote{The use of the Mathematica command {\tt Simplify} does not identify the factors $(\Phi_2(s))^j=\pi^{j/2}(1 + {\rm erf}(x))^j$ as present in the functional form
 (\ref{tb2}) of $\{f_N^{\rm G}(1;s)\}$ but rather presents them in a binomial expansion of such factors. This remains true of using {\tt FullSimplify} in the cases $N > 6$.}
to $N$ no bigger than 12. 

Denote by $\kappa_p(N,k)$ the $p$-th cumulant of $f_N^{\rm G}(k;s)$, where in particular $\kappa_1(N,k) = \mu$ is the mean
and $\kappa_2(N,k) = \sigma^2$ is the variance.
Starting from the third cumulant, define too their scaled form by
introducing 
  \begin{equation}\label{2.12} 
\gamma_{p-2} = \kappa_p(N,k)/\sigma^p;
  \end{equation}   
here $\gamma_1$  is the skewness and $\gamma_2$ the kurtosis. In 
\cite[Appendix 4]{Ek74}, a table of $\{\mu,\sigma,\gamma_1, \gamma_2,\gamma_3,\gamma_4\}$ accurate up to
and including the 7th decimal place is presented for $N=2,\dots,7$
and $k=1,\dots,\lceil N/2 \rceil$, although an adjustment is required due to a different scale\footnote{In \cite{Ek74} the GOE eigenvalue PDF has each $x_l$ replaced by $x_l/\sqrt{2}$ relative to our
(\ref{a1.1}), implying that the corresponding values of $\mu$ presented in \cite[Appendix 4]{Ek74} must be multiplied by $\sqrt{2}$ to match
our results, and similarly the values of $\sigma$.}.
 In Appendix A below, using the exact functional form obtained from our code, we extend this data
by providing the same table, now for  $N=8,\dots,12$.

\subsection{Laguerre case}\label{S2.2}
 Relevant here are the quantities
  \begin{align}\label{1.4cL}
  & \Psi^{\rm L}(\alpha;x)  := {1 \over \Gamma(\alpha)} \int_{0}^x t^{\alpha-1} e^{- t} \, dt, \quad I^{\rm L}(\alpha,\beta;x) :=  {1 \over \Gamma(\alpha)} \int_{0}^x t^{\alpha-1} e^{- t }   \Psi^{\rm L}(\beta;t) \, dt \nonumber \\
   & \tilde{\Psi}^{\rm L}(\alpha;x)  := {1 \over \Gamma(\alpha)} \int_x^{\infty} t^{\alpha - 1} e^{- t} \, dt, \quad \tilde{I}^{\rm L}(\alpha,\beta;x) :=  {1 \over \Gamma(\alpha)} \int_x^{\infty} t^{\alpha-1} e^{- t}  \tilde{\Psi}^{\rm L}(\beta;t) \, dt.
  \end{align}
  They allow (\ref{1.3}) and  (\ref{1.4b}) to be written
  \begin{align}\label{1.2aL}  
   & H_0 ^{\rm L}(j,k;s) = -   I^{\rm L}(a+j,a+k;s)  +   I^{\rm L}(a+k,a+j;s),   \nonumber \\
     & H_1^{\rm L}(j,k;s) =    -\tilde{\Psi}^{\rm L}(a+j;s)    {\Psi}^{\rm L}(a+k;s)  +    \tilde{\Psi}^{\rm L}(a+k;s)    {\Psi}^{\rm L}(a+j;s)  \nonumber \\
   &  \hspace*{1.75cm}=     -   {\Psi}^{\rm L}(a+k;s)  +        {\Psi}^{\rm L}(a+j;s) \nonumber \\
    & H_2^{\rm L}(j,k;s) =     \tilde{I}^{\rm L}(a+j,a+k;s)  -   \tilde{I}^{\rm L}(a+k,a+j;s) , \quad    \nu_j^{\rm L}(s)  =  \Psi^{\rm L}(a+j;s) + \zeta  \tilde{\Psi}^{\rm L}(a+j;s).
    \end{align}  
   Chiani's \cite{Ch14} strategy of integration by parts used in relation to $\Psi^{\rm L}$ and $ I^{\rm L}$  allows all the functions in (\ref{1.4cL}) to be determined by first order recurrences and appropriate initial conditions.
   
   \begin{prop}\label{P2.3}
   We have the recurrences
     \begin{align}\label{1.3L}  
     & \Psi^{\rm L}(\alpha;x)  = - {x^{\alpha - 1} \over \Gamma(\alpha)} e^{- x} +  \Psi^{\rm L}(\alpha - 1;x), \quad  \tilde{\Psi}^{\rm L}(\alpha;x)  =   {x^{\alpha - 1} \over \Gamma(\alpha)} e^{- x} +  \tilde{\Psi}^{\rm L}(\alpha - 1;x),    \nonumber \\
     &  I^{\rm L}(\alpha+1,\beta;x) =   I^{\rm L}(\alpha,\beta;x) -  {x^{\alpha}   e^{- x }  \over \Gamma(\alpha+1)}  \Psi^{\rm L}(\beta;x)  + {2^{-(\alpha+\beta)}  \Gamma(\alpha+\beta) \over  \Gamma(\alpha+1)  \Gamma(\beta) }
       \Psi^{\rm L}(\alpha+\beta;{2} x),    \nonumber \\
  &  \tilde{I}^{\rm L}(\alpha + 1,\beta ;x) =   \tilde{I}^{\rm L}(\alpha,\beta;x) +  {x^{\alpha}   e^{- x}  \over \Gamma(\alpha+1)}  \tilde{\Psi}^{\rm L}(\beta;x)  - {2^{-(\alpha+\beta)}  \Gamma(\alpha+\beta) \over  \Gamma(\alpha+1)  \Gamma(\beta) }
       \tilde{\Psi}^{\rm L}(\alpha +\beta; 2 x).     
     \end{align} 
     \end{prop}
     
     Using the recurrences (\ref{1.3L}), together with the symmetry relations 
      \begin{equation}\label{2.5bL} 
    I^{\rm L}(\alpha,\beta;x)   = \psi^{\rm L}(\alpha;x)  \psi^{\rm L}(\beta;x) -   I^{\rm L}(\beta,\alpha;x), \quad     \tilde{I}^{\rm L}(\alpha,k;x)   = \tilde{\psi}^{\rm L}(\alpha;x)  \tilde{\psi}^{\rm L}(\beta;x) -   \tilde{I}^{\rm L}(\beta,\alpha;x)
      \end{equation}  
      (as with (\ref{2.5b}), these follow by an appropriate integration by parts),
     the formulas of (\ref{1.2aL}) imply recurrences for $H_0^{\rm L}$ and $H_2^{\rm L}$, the first of which can be found in \cite[Eq.~(18)]{Ch14}.
     
     \begin{cor}
     We have
    \begin{multline}\label{2.6aL}     
H_0^{\rm L}(j,k+1;s) = H_0^{\rm L}(j,k;s) - {x^{a+k} \over \Gamma(a+k+1)} e^{-x} \Psi^{\rm L}(a+j;x)  \\  + 2^{-(2a+j+k-1)}
{\Gamma(2a+j+k) \over \Gamma(a+k+1) \Gamma(a+j)}  \Psi^{\rm L}(2a+j+k;2x) 
 \end{multline} 
 and
  \begin{multline}\label{2.6bL}     
{H}_2^{\rm L}(j,k+1;s) = {H}_2^{\rm L}(j,k;s) - {x^{a+k} \over \Gamma(a+k+1)} e^{-x} \tilde{\Psi}^{\rm L}(a+j;x)  \\  + 2^{-(2a+j+k-1)}
{\Gamma(2a+j+k) \over \Gamma(a+k+1) \Gamma(a+j)}  \tilde{\Psi}^{\rm L}(2a+j+k;2x). 
 \end{multline}  
 \end{cor}
 
 To compute $\{ H_\mu(j,k;s) \}_{j=k}^N$ ($\mu=0,1,2$) we see from (\ref{2.6aL}), (\ref{2.6bL}) and the formula for $H_1$ in (\ref{1.2aL}) that it suffices to
 have knowledge of $\{ \Psi^{\rm L}(a+j;x), \tilde{\Psi}^{\rm L}(a+j;x) \}_{j=1}^N$ and $\{ \Psi^{\rm L}(2a+j;x), \tilde{\Psi}^{\rm L}(2a+j;x) \}_{j=1}^{2N-1}$ 
 (note that the initial conditions for the recurrences  (\ref{2.6aL}) and (\ref{2.6bL}) are $H_0^{\rm L}(j,j;s) = H_2^{\rm L}(j,j;s) = 0$). If the aim is to obtain
 exact functional forms involving a minimal basis in the sense of (\ref{tb3}) and (\ref{tb4}), these functions are to be computed using the recurrences on
 the first line of Proposition \ref{P2.3}, together with the initial conditions
 \begin{equation}\label{1.3aL} 
     \Psi^{\rm L}(0;x)  =     1,      \quad   \Psi^{\rm L}(1/2;x)  =     {\rm erf} (\sqrt{x}) , \quad      \tilde{\Psi}^{\rm L}(0;x)  =     0,      \quad    \tilde{\Psi}^{\rm L}(1/2;x)  =     \Big ( 1 - {\rm erf} (\sqrt{x}) \Big ).  
      \end{equation} 
The latter are relevant to  the setting of $ \mathcal P^{\rm L}$ in (\ref{a1.2}) that the parameter $a$ is a non-negative integer, or a half integer no less than $-1/2$.
But if the aim is numerical evaluation using Mathematica, the facts that
 \begin{equation}\label{1.7L}
 \Psi^{\rm L}(\alpha;x) = 1 -  { \Gamma(\alpha;x) \over \Gamma(\alpha)}, \quad \tilde{\Psi}^{\rm L}(\alpha;x) =  {\Gamma(\alpha;x) \over \Gamma(\alpha)}, 
\end{equation} 
where $ \Gamma(\alpha;x) := \int_x^\infty t^{\alpha-1} e^{-t} \, dt$ is the incomplete gamma function, implies that there is no need for the
use of a recurrence, as the special function $ \Gamma(\alpha;x)$ is part of the package. Note too that $a$ in   (\ref{a1.2}) can take continuous
values from this viewpoint.

 \section{Pfaffian formulation and numerical evaluations}
 The use of the symbol $\zeta$ in (\ref{1.0}) can be replaced by the use of  complex roots of unity according to the Fourier sum formula
  \begin{multline}\label{3.1}
  (N + 1) E_N^{(\cdot)}(k;(s,\infty))  =  \sum_{l=0}^N e^{-2 \pi i k l /(N+1)} \Xi_N^{(\cdot)}((s,\infty);   e^{2 \pi i  l /(N+1)}) \\
   =   \begin{cases} (1 + 2 {\rm Re} \, \sum_{l=1}^{N/2}  e^{-2 \pi i k l /(N+1)} \Xi_N^{(\cdot)}((s,\infty);   e^{2 \pi i  l /(N+1)}), & N \: {\rm even} \\
  1 + (-1)^k  \Xi_N^{(\cdot)}((s,\infty);   -1) + 2 {\rm Re} \, \sum_{l=1}^{(N-1)/2}  e^{-2 \pi i k l /(N+1)} \Xi_N^{(\cdot)}((s,\infty);   e^{2 \pi i  l /(N+1)}) , & N \: {\rm odd}, 
  \end{cases}
  \end{multline} 
  where to obtain the second line use has been made of the normalisation requirement that $ \Xi_N^{(\cdot)}((s,\infty);   1) = 1$.
The significance of this is that with $\zeta$ a specific (complex) value, for a given value of $s$, the matrix elements in the Pfaffian formulas for $  \Xi_N^{(\cdot)}$ in Proposition \ref{P1} can all be 
calculated numerically rather than symbolically, thus allowing for a numerical determination of $E_N^{(\cdot)}(k;(s,\infty))$. 

\begin{figure*}
\centering
\includegraphics[width=1.0\textwidth]{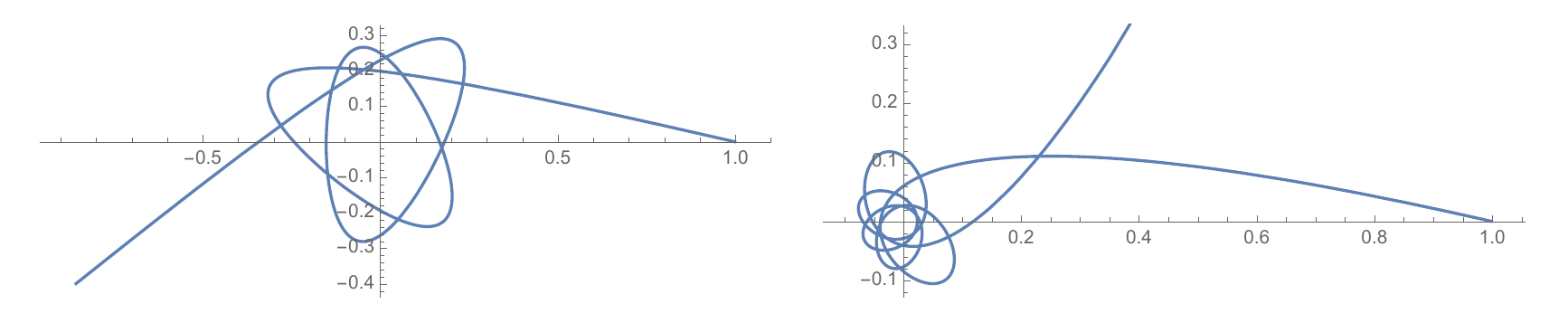}
\caption{Plotted parametrically are the values in the complex plane of  $ \Xi_N^{\rm G}((s,\infty);   e^{2 \pi i  l /(N+1)})$ (left panel)
and  $(\Xi_N^{\rm G}((s,\infty);   e^{2 \pi i  l /(N+1)}))^2$ (right panel) for $N=6$ and $l=3$ The parameter is the variables $s$, which is taken in the range $s=5$
(at this value of $s$  the curve (graphically) is at the real value 1 in the complex plane) and decreased down to $s=-4$ (which corresponds to the other end of the curve).}
\label{F1}
\end{figure*}

For $A$ a general $2n \times 2n$
  anti-symmetric matrix, there is the standard result (see e.g.~\cite[Eq.~(6.12)]{Fo10}) that
  \begin{equation}\label{3.2}  
  ( {\rm Pf}(A))^2 = \det(A).
  \end{equation}
  In the case that $A$ has numerical entries and one has prior knowledge that the Pfaffian is positive, by taking the square root this formula
  can be used to provide an efficient computation of $ {\rm Pf}(A)$. However, when the Pfaffian is a general complex number as in (\ref{3.1})
  for $l \ne 0$, a decision has to be made about the correct branch of the square root. For the present setting, by noting from (\ref{1.0}) that for large $s$ 
  we have $ \Xi_N^{(\cdot)}((s,\infty), \zeta) \sim E_N^{(\cdot)}(0;(s,\infty)) \sim 1$ independent of $\zeta$, a possible way to proceed using (\ref{3.2})
  would be to start computing
  $ \Xi_N^{(\cdot)}((s,\infty), \zeta)$ for a fixed complex $\zeta$ with $s$ large, for which its value is approximately 1, and then to reduce $s$ in small intervals.
  The branch of the square root required by (\ref{3.2}) is to be chosen by the requirement of approximate (with respect to the small interval size) continuity in values.
  In Figure \ref{F1} this task is illustrated in a particular Gaussian case with $N=6, l=3$,
  by comparing the values in the complex plane of $ \Xi_N^{\rm G}((s,\infty);   e^{2 \pi i  l /(N+1)})$  
  obtained by decreasing $s$ from $s=5$ to $s=-4$ with the values of $(\Xi_N^{\rm G}((s,\infty);   e^{2 \pi i  l /(N+1)}))^2$. Increasing the size of $N$, analogous
  plots are observed but with more and tighter rotations about the origin, making the task of choosing the correct branch of the square root for discrete increments
  in $s$ a delicate task.
  
  Fortunately, since the 2012 work on Wimmer \cite{Wi12} and in particular the software provided as part of the corresponding arXiv posting, it is now
  possible to efficiently compute ${\rm Pf}(A)$ for $A$ a general $2n \times 2n$
  anti-symmetric matrix with numerical entries directly, independent of  (\ref{3.2}) and the associated square root issue. This is based on unitary conjugations reducing
  $A$ to a skew symmetric tridiagonal form. For given $N,l,s$ it is the Mathematica code  by Wimmer, applied to the Pfaffian formulas of
  Proposition \ref{P1}, which we use to compute $ \Xi_N^{(\cdot)}((s,\infty);   e^{2 \pi i  l /(N+1)})$
  in (\ref{3.1}).
  
  There is an important point to make in relation to the loss of conditioning with respect to increasing values of $N$. This point is that the computation
  of determinants and Pfaffians is, in general, ill-conditioned with respect to truncations of the values of the entries \cite{St73}. Thus it is necessary to increase
  the number of digits in the floating point arithmetic with $N$, which is simple to do using Mathematica.

\subsection{Gaussian case}\label{S3.1}
We will consider first the sequence of conditioned gap probabilities $\{ E_N^{\rm G}(k;(s,\infty)) \}_{k=0}^N \}$ determining
the random variable  ${\mathcal N}_{(s,\infty)} := \sum_{l=1}^N \chi_{x_l \in (s,\infty)}$
for the number of eigenvalues in the semi-infinite interval $(s,\infty))$. The specific case $s=0$ corresponds to
the number of positive eigenvalues, which has attracted particular interest  for its relevance to stability questions in disordered
systems \cite{CGG00}, landscape based string theory \cite{AE06}, quantum cosmology \cite{MH06}, among other examples;
see the introduction to \cite{MNSV11} for more references.

On the basis of some approximate analysis, it was predicted in  \cite{CGG00}  that in a neighbourhood of the mean $\mu_N:= \langle
{\mathcal N}_{(0,\infty)} \rangle = N/2$, and for large $N$, the random variable ${\mathcal N}_{(0,\infty)}$ satisfies the local central limit
theorem
 \begin{equation}\label{3.3}  
 {\rm Pr} \Big (({\mathcal N}_{(0,\infty)} - \mu_N )= k \Big )  \to {1 \over \sqrt{2 \pi \sigma_N^2}}  e^{-k^2/(2\sigma_N^2)}, \quad \sigma_N^2 = 
  {1 \over \pi^2} \log N;
  \end{equation}
  see also \cite{MNSV11}. Note that for this to be a meaningful limit law one must have $k/\sigma_N$ to be of order unity as $N \to \infty$; also, for convenience, in (\ref{3.3}) it is assumed that $N$ is even. A rigorous determination of the leading order variance $\sigma_N^2$ can be found in \cite{Sh15}
  (see the review \cite[Remark 3.1.3]{Fo23} for related references).
  Our computational scheme applied to $\{ E_N^{\rm G}(k;(0,\infty)) \}_{k=0}^N \}$ can test the prediction (\ref{3.3}) and moreover probe correction terms in
  a large $N$ expansion. 
  
  Before doing so, some remarks along these lines in relation to the random variable  ${\mathcal N}_{(0,\infty)}$ for the Gaussian unitary ensemble (GUE) of complex Hermitian
  matrices (see e..g.~\cite[\S 1.3.1]{Fo10}) are in order. Up to the scaling $x_l \mapsto \sqrt{2} x_l$, the eigenvalue PDF for the GUE is given by (\ref{a1.1}) with the product
  therein now squared. Whereas the eigenvalue PDF for the GOE is an example of a Pfaffian point process, the  eigenvalue PDF for the GUE is an example of a determinantal
  point process, which is a simpler structure with additional integrability properties. Leveraging the latter, it was proved in \cite[Eq.~(143)]{WF12} (see also \cite{MNSV11}
  for the first two leading orders)
  \begin{multline}\label{3.4}   
  {\rm Var} \Big ({\mathcal N}_{(0,\infty)}^{\rm GUE} \Big ) = {\log 4 N + \gamma + 1 \over 2 \pi^2} - {\log 4 N + \gamma \over 4 \pi^2 N} + {7 \over 96 \pi^2 N^2} +
  {24 \log 4 N + 24 \gamma - 41 \over 192 \pi^2 N^3} \\
  - {219 \over 5120 \pi^2 N^4 } + {\rm O}(N^{-5} \log N),
   \end{multline}
    where $\gamma$ denotes Euler's constant and for convenience $N$ is assumed even (the analogue for $N$ odd was also derived, as were the explicit form of the terms up to
    $ {\rm O}(N^{-7} \log N)$). The determinantal structure implies that the generating function $\Xi_N^{\rm GUE}((0,\infty);\zeta)$, defined as in (\ref{1.0}), has all its zeros on the
    negative real axis in the complex $\zeta$-plane. This, combined with knowledge of a central limit theorem for ${\mathcal N}^{\rm GUE}_{(0,\infty)}$ \cite{CL95,So00}, implies
    the validity of the local  central limit theorem (\ref{3.3}) in the GUE case, now with $\sigma_N^2 =   {1 \over 2 \pi^2} \log N$ \cite{FL14}.
    
    Returning now to the consideration of $\{ E_N^{\rm G}(k;(0,\infty)) \}_{k=0}^N \}$, we first list in Table \ref{TA} our computed values of 
 \begin{equation}\label{3.5} 
  {\rm Var} \Big ({\mathcal N}_{(0,\infty)}^{\rm GOE} \Big ) :=     \sum_{k=0}^N (k - N/2)^2  E_N^{\rm G}(k;(0,\infty))
  \end{equation}
  for $N$ starting at 10 and finishing at 100, with increments of 10. 
  
  \vspace{.5cm}
  
 \begin{table}[ht]
  \centering
  \begin{tabular}{ll}
    \begin{tabular}{c|c}
$N$ &    $ {\rm Var}  ({\mathcal N}_{(0,\infty)}^{\rm GOE}  )$   \\
    \hline
    10 & 0.4735719613 \\
    20 & 0.5464313214  \\
    30  &0.5883873972 \\ 
   40 & 0.6179704941  \\
   50 &  0.6408395677
    \end{tabular}\hfill
    &
    \begin{tabular}{c|c}
$N$ &    $ {\rm Var}  ({\mathcal N}_{(0,\infty)}^{\rm GOE}  )$   \\
    \hline
    60 & 0.6594853769 \\
    70 &0.6752272387   \\  
    80  & 0.6888489801  \\  
   90 &  0.7008544880  \\  
   100 &  0.7115869419
    \end{tabular}
     \end{tabular}
      \caption{10 decimal place accurate variances (\ref{3.5})}
         \label{TA}
 \end{table}%
 
 Guided by (\ref{3.4}) and knowledge of the leading term as given by $\sigma_N^2$ in (\ref{3.3}), we make the ansatz
  \begin{equation}\label{3.5a}   
    \pi^2   {\rm Var}  ({\mathcal N}_{(0,\infty)}^{\rm GOE}  ) - \log N = {c_1} + {c_2 \log N \over N} + {c_3 \over N} + \cdots
    \end{equation}
    to fit to the data of Table \ref{TA}. We do this by choosing various combinations of 3 rows of the tables, which results in the values
    $c_1 \approx 2.4229$, $c_2 \approx 0.006$, $c_3 \approx -0.52$. The small value of $c_2$ relative to $c_3$ puts in doubt 
   the correctness of (\ref{3.5a}), which may then be distinct to (\ref{3.4}) in the second leading term. On the other hand, the value of $c_1$ is available in the literature as $3 \log 2 + \gamma + 1 - {\pi^2 \over 8} = 2.42295\dots$ \cite[Eq.~(40) with $a=0$, $\beta =1$]{SLMS21}, thus providing evidence for the accuracy of our numerical values.
   
   Next we make some remarks relating to correction terms to the local central limit law (\ref{3.3}). As already remarked, the appropriate scaling
   variable is $k/\sigma_N$. But with $k$ taking on integer values only, and $\sigma_N = {1 \over \pi } (\log N)^{1/2}$, numerical tabulation for moderate
   values of $N$ (say between 50 and 100) are aways probing only the tails of the limiting distribution. Even a value of $N=1000$ (which is out of reach for
   the practical implementation of our numerical methods due to the ill-conditioning), the value of $\sigma_N$ is $0.91...$, which implies varying $k$ by 1
   is more than one unit of standard deviation.\footnote{The problem of the distribution of the length of the longest increasing subsequence of a random
   permutation, which has well known relationships to random matrix theory \cite{BDJ98,BF03}. Here $\sigma_N \sim N^{1/6}$, and is an example where large
   $N$ data has successfully been generated for the numerical determination of the leading correction \cite{FM23,Bo23}, leading in turn to its analytic
   determination \cite{Bo24}. Another example is in relation to the local central limit theorem satisfied by the real zeros of elliptic GinOE matrices \cite{Fo24};
   see the recent book \cite{BF23} for more on this class of random matrix ensemble.}
   The expansion for the variance (\ref{3.5}) suggests the scaled large $N$ expansion
     \begin{equation}\label{3.6} 
   {\rm Pr} \Big (({\mathcal N}_{(0,\infty)}^{\rm GOE}  - \mu_N) = k \Big )  \sim {1 \over \sqrt{2 \pi  \sigma_N^2}}  e^{-k^2/(2 \sigma_N^2)} + {1 \over \sigma_N^2} 
   P_1(k/\sigma_N) + \cdots
   \end{equation} 
   for some functional form $P_1$. A question in keeping with recent literature \cite{FT19a,Bo24a,BL24} relates to (\ref{3.6}) identifying the optimal rate of convergence,
   meaning can $\sigma_N^2$ herein be replace by a function of $N$ with leading large $N$ form  $\sigma_N^2$ and with the effect of eliminating the correction term
   in (\ref{3.6})? This will happen when $P_1$ is related to the leading term by a derivative operation \cite{PS16,FT18}.
   
As far as illustrating (\ref{3.3}) goes (where we replace $\sigma_N^2$ by $ {\rm Var}  ({\mathcal N}_{(0,\infty)}^{\rm GOE})$ as is usual in stating a local
central limit theorem), 
 in Table \ref{TB}, where we compare $p_N^{\rm exact}(k) :=   {\rm Pr}  ({\mathcal N}_{(0,\infty)}^{\rm GOE}  - \mu_N=k)$
   against 
    \begin{equation}\label{3.6b}   p_N^{\rm approx}(k) :=  {1 \over \sqrt{2 \pi  {\rm Var}  ({\mathcal N}_{(0,\infty)}^{\rm GOE}  )}}  e^{-k^2/(2  {\rm Var}  ({\mathcal N}_{(0,\infty)}^{\rm GOE}  ))}.
     \end{equation}  
 We remark that a theoretical justification of such a relation can, following \cite{FL14}, be undertaken by knowledge of the location of the zeros of the generating function (\ref{1.0}); see the numerically evidenced discussion in \cite[\S 3.5]{FK24}.

  \vspace{.5cm}
  
 \begin{table}[ht]
  \centering
  \begin{tabular}{ll}
    \begin{tabular}{c|c|c|c}
$k$ &    $p_{70}^{\rm exact}(k) $ &  $p_{70}^{\rm approx}(k)$ & $\delta_{70}(k) $ \\
    \hline
    0 &  0.4838115 & 0.4854953 & $-$0.0016838\\
    1 &   0.2325255 &  0.2315227 & 0.0010028 \\
    2   & 0.0250092 & 0.0251084  & $-$0.0000992 \\ 
   3   & 0.0005570 & 0.0006192 &  $-$0.0000622 \\
    \end{tabular}\hfill
    &
    \begin{tabular}{c|c|c|c}
$k$ &    $p_{100}^{\rm exact}(k) $ &  $p_{100}^{\rm approx}(k)$ & $\delta_{100}(k) $ \\
    \hline
    0 & 0.4714611  & 0.4729291 & $-$0.0014681\\
    1 &   0.2350801 &  0.2342270 &   0.0008531 \\
    2   &  0.0284044 & 0.0284550  & $-$0.0000506 \\ 
   3   &  0.0007803 & 0.0008479 &  $-$0.0000676 \\
     \end{tabular}
        \end{tabular}
      \caption{Here $\delta_N(k):= p_N^{\rm exact}(k)  -  p_N^{\rm approx}(k)$ }
         \label{TB}
 \end{table}%
 
In applications to the stability questions cited in the beginning paragraph of this section, there is specific
interest in the large $N$ form of $ {\rm Pr}  ({\mathcal N}_{(0,\infty)}  = 0  )$, which is in the large
deviation regime with respect to (\ref{3.3}). For the GUE, it has been proved \cite{BEMN11, DS17} that
\begin{equation}
\log  {\rm Pr} \Big ({\mathcal N}_{(0,\infty)}^{\rm GUE}  = 0 \Big ) =  c_1 N^2 + c_2 \log N + c_3 + \cdots
\end{equation}
with $c_1 = - {1 \over 2} \log 3$, $c_2 = - {1 \over 12}$, $c_3 = {1 \over 8} \log 3 - {1 \over 6} \log 2 + \zeta'(-1)$.
In fact the results of \cite{BEMN11} apply to the GOE case well, telling us that
\begin{equation}\label{3.6x}
\log  {\rm Pr} \Big ({\mathcal N}_{(0,\infty)}^{\rm GOE}  = 0 \Big ) = \tilde{c}_1 N^2 + \tilde{c}_2  N +  \tilde{c}_3 \log N + \tilde{c}_4 + \cdots
\end{equation}
with $\tilde{c}_1 = -{1 \over 4} \log 3$, $\tilde{c}_2 = - {1 \over 2} \log (1 + {2 \over \sqrt{3}})$, $\tilde{c}_3 =- {1 \over 24}$,
$\tilde{c}_4 = -{1 \over 12} \log 2 - {1 \over 16} \log 3 + {1 \over 4} \log (
   1 + {2 \over \sqrt{3}} )+ {1 \over 2} \zeta'(1)$. Our high precision evaluation of $E_N^{\rm G}(0;(0,\infty))$ (which
   by definition is equal to   ${\rm Pr} ({\mathcal N}_{(0,\infty)}^{\rm GOE}  = 0  ) $) for various $N$ is used in Table
   \ref{T4} to illustrate
   the accuracy of the asymptotic expansion (\ref{3.6x}).
   
  \vspace{.5cm}
  
 \begin{table}[ht]
  \centering
    \begin{tabular}{c|c|c|c}   
 $N$ & $\log E_N^{\rm G}(0;(0,\infty))$ & $\log \tilde{E}_N^{\rm G}(0;(0,\infty))$ & $\delta_N$ \\
    \hline
    10 &  $-$31.4183282    & $-$31.4167301 & $-$0.0015980 \\
    20 & $-$117.6805735    & $-$117.6797917 & $-$0.0007817 \\
   30 &  $-$258.8619980   & $-$258.8614809 & $-$0.0005170
    \end{tabular}
      \caption{Here $\log \tilde{E}_N^{\rm G}(0;(0,\infty))$ is defined as the asymptotic expansion on the right hand side of
      (\ref{3.6x}), and $\delta_N:= \log {E}_N^{\rm G}(0;(0,\infty)) - \log \tilde{E}_N^{\rm G}(0;(0,\infty))$. Notice that
      $\delta_N$ is decreasing at the rate $1/N$ as $N$ increases.}
      \label{T4}
 \end{table}      
 
 We turn our attention now to the marginal eigenvalue PDFs in the bulk, choosing for
 simplicity $f_N^{\rm G}((N+1)/2;s)$ with $N$ odd, this (by symmetry) being an even function of
 $s$. We use our computational scheme to compute the sum in (\ref{a1.3}) for discrete values
 of $s$, small with respect to the scale of $\sigma_N$. An interpolating function connecting these values
 is then formed within the Mathematica software, which allows for the derivative operation required in (\ref{a1.3})  to be carried out.
 As mentioned in the Introduction, there are limit laws available for the individual eigenvalues, moving inwards
 from the edge with the matrix size. Such limit laws were first derived in the GUE case by Gustavsson \cite{Gu05}, with
 the main tool being the determinantal structure and its knowledge of a central limit theorem for the counting statistic
 ${\mathcal N}_{(s,\infty)}$ from \cite{So00}. Subsequently, it was realised by O'Rourke \cite{OR10} that the use of the
 inter-relation even(GOE${}_N \cup{\rm GOE}_{N+1}$) = GUE${}_N$ from \cite{FR01} (here the operation 
 GOE${}_N \cup{\rm GOE}_{N+1}$ denotes the random superposition of the spectrum of a GOE ensemble with $N$ eigenvalues,
 and a GOE ensemble of $N+1$ eigenvalues, while the operation even$(\cdot)$ denotes observing only the even labelled
 eigenvalues when reading from either edge with the eigenvalues labelled successively) allows the results of \cite{Gu05} to be extended to the GOE case.
 Specifically, for a GOE eigenvalue $x_k$ near or at the centre of the spectrum (for us $x_{(N-1)/2}$ with $N$ odd), we have from \cite[Remark 8]{OR10} that
 \begin{equation}\label{3.8}
\lim_{N \to \infty}   \Big ( {2N \over \log N } \Big )^{1/2} x_k  \mathop{=}\limits^{\rm d} {\tt N}[0,1],
\end{equation}
where $ {\tt N}[0,1]$ denotes the standard normal distribution. 
(An alternative derivation of this result, extended to the Gaussian $\beta$ ensemble, has recently been given in \cite{FTW23}.)
One remarks that in this formula the factor of $\sqrt{N}$ can be interpreted as resulting from the value of the density of the GOE at the origin, which to leading order is a scaled semi-circle; see \cite[Eq.~(1.52)]{Fo10}.

\begin{figure*}
\centering
\includegraphics[width=0.7\textwidth]{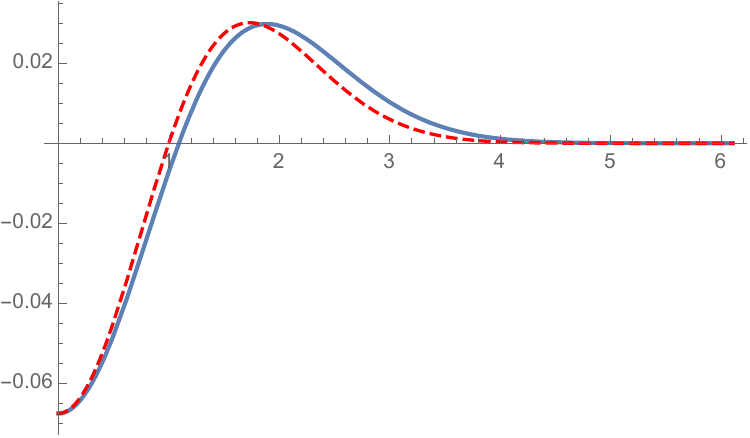}
\caption{[color online] Plot of the difference (\ref{3.10}) for $N=41$ (blue solid curve), superimposed with the approximation (\ref{3.11}) (red dashed curve) for $c$ chosen for a matching
at the origin. The graphs are shown for $X \ge 0$ only, as both are symmetrical about the origin. 
}
\label{F2}
\end{figure*}

In keeping with (\ref{3.8}), introduce the scaled eigenvalues 
 \begin{equation}\label{3.9u}
X_k =  \Big ( {2N \over \log N } \Big )^{1/2} x_k.
\end{equation}
 Denote by $f_N^{\rm G}(k;X)$ the PDF of $X_k$,
as is consistent with the notation used in (\ref{a1.3i}).
Our interest is in using our ability to compute this PDF, with $k=(N-1)/2$ for a sequence of $N$ values, to probe the leading rate of convergence
to the limit law (\ref{3.8}). Thus we seek the function of $N$, $\alpha(N)$ say, with $\alpha(N) \to 0$ as $N \to \infty$, and the function of $X$, $h(X)$ say,
such that for large $N$ one has the asymptotic expansion
 \begin{equation}\label{3.9}
 \Big ( {2N \over \log N } \Big )^{1/2} f_N^{\rm G}((N-1)/2;X) \sim {1 \over \sqrt{2 \pi}} e^{-X^2/2}  + \alpha(N) h(X) + \cdots,
\end{equation}
where terms not written decay at a rate fast than $\alpha(N)$. A numerical approach can access the difference 
 \begin{equation}\label{3.10}
 \Big ( {2N \over \log N } \Big )^{1/2}  f_N^{\rm G}((N-1)/2;X)  - {1 \over \sqrt{2 \pi}} e^{-X^2/2},
\end{equation}
 which is
only an approximation to $ \alpha(N) h(X) $ as it contains all the higher order terms in the expansion (\ref{3.9}) as well. Nonetheless, one is lead to the prediction that $\alpha(N) =
{1 \over \log N}$, and to (an approximation of) the graphical form of $h(X)$ after computing (\ref{3.10}) for just the two values of $N$, $N=21$ and $N=41$. Moreover, inspection of the
graphical form shows that it closely resembles (but is not equal to)
 \begin{equation}\label{3.11}
c \Big (1 - X {d \over d X} \Big )  e^{-X^2/2}
\end{equation}
for a certain constant $c$; see Figure \ref{F2}.

Suppose now that instead of the scaling (\ref{3.9u}) one was to introduce $\tilde{X}_k :=  \Big ( { \log N \over 2N} \Big )^{1/2}(1 + \alpha(N)) x_k$. Taylor expanding with respect to
the small variable $\alpha(N)$ to reclaim the variable (\ref{3.9u}) shows
 \begin{multline}\label{3.10u}
 \Big ( { 2N \over \log N} \Big )^{1/2} (1 + \tilde{c} \alpha(N))  f_N^{\rm G}((N-1)/2;\tilde{X})    \\
 \sim {1 \over \sqrt{2 \pi}} e^{-X^2/2} +   \alpha(N) \bigg (  \tilde{c}   \Big (1 - X {d \over d X} \Big )  e^{-X^2/2} +
   h(X) \bigg ) + \cdots,
\end{multline}
Thus, if it were to be that (\ref{3.11}) was proportional to $h(X)$ it would be possible to choose $\tilde{c}$ to improve the rate
of convergence by eliminating the term proportional to $\alpha(N)$; see \cite{PS16,FT19a} for some examples. However, in the present
situation, while the functional forms are very similar, they are not exactly the same, so such an improvement is not possible. On the other
hand, from the viewpoint of numerical values rather than the rate of convergence, it follows that the PDF for the scaled eigenvalue
$ X_{(N-1)/2}/{\rm Var}(X_{(N-1)/2})$, rather than for (\ref{3.9u}), more accurately follows $ {\tt N}[0,1]$ in distribution for finite $N$ (cf.~(\ref{3.6b})).

\begin{remark}
A question of much interest (see e.g.~\cite{BFM17} for motivation), but not accessible via our present results, is the finite size corrections for the scaled spacing distribution of two GOE eigenvalues
near the centre of the spectrum (say $x_{N/2}$ and $x_{(N/2) + 1}$ for $N$ even). In the case of the circular version of the GOE --- the circular orthogonal ensemble (COE) 
(see \cite[\S 2.2.2]{Fo10}) --- the leading correction term to the large $N$ limit law is proportional to $1/N^2$ \cite{FLT20}.
\end{remark}

\subsection{Laguerre case}\label{S3.2}
Relevant to the consideration of large $N$ limit laws in the Laguerre case is the limiting eigenvalue density (the limiting eigenvalue density is also relevant in the Gaussian case, but didn't appear in our discussion since our consideration of the bulk was restricted to the neighbourhood of the origin for simplicity of presentation --- in the Laguerre case such a simplification is not possible as the spectrum has no such point of symmetry). First, scale the LOE eigenvalues by writing $\Lambda_j = \lambda_j/N$. Set ${n \over N} = {1 \over c}$, where $n \ge N$ is defined as in (\ref{a1.0}). The density of $\{ \Lambda_j \}$, normalised to integrate to unity, is given by the Marchenko-Pastur law \cite{PS11}
 \begin{equation}\label{MP1}
 \rho^{\rm MP}(x) = {1 \over 2 \pi} {\sqrt{(c_+ - x)(x - c_-)} \over c x} \mathbbm 1_{x \in (c_-,c_+)},
 \end{equation}
 with $c_{\pm} = (1 \pm \sqrt{c})^2$. By definition of the density, for large $N$ the expected number of the scaled eigenvalues in the interval $(\tilde{s},c_+)$ is $N \mu_{\tilde{s}}$, $\mu_{\tilde{s}} = 
 \int_{\tilde{s}}^{c_+} \rho^{\rm MP}(x) \, dx$. With $\mathcal N_{(\tilde{s},c_+)}^{\rm L}(\{\Lambda_j\})$ the random variable for the number of scaled eigenvalues in $(\tilde{s},c_+)$, considerations from the topic of principal component analysis motivated a study in \cite{MV12} of the fluctuation ${\rm Var} \, \mathcal N_{(\tilde{s},c_+)}^{\rm L}(\{\Lambda_j\})$. This quantity was shown to exhibit the asymptotic form ${1 \over \pi^2} \log N$ (which is the same as $\sigma_N^2$ in (\ref{3.3})) independent of the value of $\tilde{s}$ ($c_- < \tilde{s} < c_+$), and moreover it was argued that a local central limit theorem quantitatively the same as (\ref{3.3}) holds true.

Our ability to compute $\{E_N^{\rm L}(k;(s,\infty))\}_{k=0}^N$ from \S \ref{S2.2} allows for the above predictions to be illustrated and checked. First we consider
\begin{equation}\label{E1}
\Big \langle \mathcal N_{(\tilde{s},\infty)}(\{\Lambda_j\} \Big \rangle =
\sum_{k=1}^\infty k E_N^{\rm L}(k;(s,\infty))
\end{equation}
in the specific case that the Laguerre parameter $a$ in (\ref{a1.2}) is equal to $1$ and $\tilde{s} = 1$ (which implies $\tilde{\mu} = 0.391\dots$). in Table \ref{T4a} we compare our high precision numerical evaluations of (\ref{E1}) against the leading order theoretical value $N \tilde{\mu}$, for a sequence of $N$ values, tabulating too their difference. The latter shows evidence of converging to a constant value at a rate proportional to $1/N$.

\begin{table}[ht]
  \centering
    \begin{tabular}{c|c|c|c}
$N$ &    $ \Big \langle \mathcal N_{(\tilde{s},\infty)}(\{\Lambda_j\} \Big \rangle$  & $N \tilde{\mu} $ & $\delta_N $ \\
    \hline
    20 & 8.711490  & 7.8200 & 0.8914 \\
    40  & 16.543998 & 15.6401 & 0.9039 \\
   60 & 24.368259  & 23.4601 & 0.9081 \\
   80 &  32.190428 & 31.2802&0.9102
    \end{tabular}
    \caption{8 digit  accurate means  (\ref{E1}), their leading order theoretical value $N \tilde{\mu}$, and the difference $\delta_N$. }
         \label{T4a}
    \end{table}

For definiteness we take the Laguerre parameter $a$ in (\ref{a1.2}) equal to $1$, which implies $c=1,c_-=0$ and $c_+=4$, and we take $s=N$ (i.e.~$\tilde{s} = 1$ which implies $\mu_{\tilde{s}} = {2 \over \pi} \int^1_{1/4}x^{-1/2}(1-x)^{1/2} \, dx$. The variance can be computed from the formula
\begin{equation}\label{3.5L} 
  {\rm Var} \Big ({\mathcal N}_{(\tilde{s},c_+)}^{\rm L}(\{ \Lambda_j \} \Big ) :=     \sum_{k=0}^N (k -  M_{\tilde{s}})^2  E_N^{\rm L}(k;(N \tilde{s},\infty)), \quad 
  M_{\tilde{s}} := \Big \langle \mathcal N_{(\tilde{s},\infty)}(\{\Lambda_j\} \Big \rangle
  \end{equation}
  (cf.~(\ref{3.5})), which we do in Table \ref{TA1} for the same parameters as in Table \ref{T4a}. We compare these values against the leading order theoretical prediction  
  ${1 \over \pi^2} \log N$, and display too the difference, which shows evidence of tending to a constant.
  
  \begin{table}[ht]
  \centering
    \begin{tabular}{c|c|c|c}
$N$ &    $ {\rm Var}  ({\mathcal N}_{(\tilde{s},c_+)}^{\rm L}  )$ & ${1 \over \pi^2} \log N$ & $\delta_N$ \\
    \hline
    20 & 0.5096148748  & 0.3035 & 0.2060 \\ 
   40 & 0.5779495211 &0.3737 & 0.2041  \\
    60 &  0.6182432846 & 0.4148 & 0.2033\\  
    80  & 0.6469688861 & 0.4439 & 0.2029 \\  
     \end{tabular}
      \caption{10 decimal place accurate variances (\ref{3.5}) against their leading order theoretical value, and the difference $\delta_N$}
         \label{TA1}
 \end{table}%

It is easy to check from the definitions that ${\rm Var} \Big ({\mathcal N}_{(\tilde{s},c_+)}^{\rm L}(\{ \Lambda_j \} \Big ) =
{\rm Var} \Big ({\mathcal N}_{(c_-,\tilde{s})}^{\rm L}(\{ \Lambda_j \} \Big )$. The significance of this is the prediction
 that in the case $c_-=0$ \cite[Eq.~(48) with $\beta=1$, $\tilde{a} =\sqrt{\tilde{s}/4}$,
$\tilde{\gamma}=0$]{SLMS21},
\begin{equation}
\lim_{N \to \infty}
\Big ({\rm Var} \Big ({\mathcal N}_{(c_-,\tilde{s})}^{\rm L}(\{ \Lambda_j \} \Big ) - {1 \over \pi^2} \log N \Big )=
{1 \over \pi^2} \Big ( 3 \log 2 + \gamma -{\pi^2 \over 8} +1 + \log( \tilde{s}^{1/2}(1 - \tilde{s}/4)^{3/2})
\Big ).
\end{equation}
Substituting $\tilde{s}=1$ gives the numerical value $0.20177\dots$ which is consistent with the values of $\delta_N$ in Table \ref{TA1}.
 
  In addition we have carried out numerical computations comparing  
\begin{equation}\label{3.6aL}  
 p_N^{\rm exact,L}(k) := {\rm Pr}\Big ( ({\mathcal N}_{(\tilde{s},c_+)}^{\rm L}(\{ \Lambda_j \} - \lfloor 
 \langle {\mathcal N}_{(\tilde{s},c_+)}^{\rm L}(\{ \Lambda_j \} \rangle 
 \rfloor)
  = k \Big )
  \end{equation}
  against the prediction from the validity of a local central limit theorem
  \begin{equation}\label{3.6bL}   p_N^{\rm approx, L}(k) :=  {1 \over \sqrt{2 \pi  {\rm Var} ( {\mathcal N}_{(\tilde{s},c_+)}^{\rm L}(\{ \Lambda_j \})}}  e^{-k^2/(2  {\rm Var}  ({\mathcal N}_{(\tilde{s},c_+)}^{\rm L}(\{ \Lambda_j \})}.
     \end{equation}
     Note that both these quantities are functions of $\tilde{s}$, which has been suppressed in our notation.
Some results of our computations, which were carried out with $\tilde{s}=1$, are presented in Table \ref{TB1}. The accuracy of (\ref{3.6bL}) is evident.

\begin{table}[ht]
  \centering
  \begin{tabular}{ll}
    \begin{tabular}{c|c|c|c}
$k$ &    $p_{60}^{\rm exact,L}(k) $ &  $p_{60}^{\rm approx,L}(k)$ & $\delta_{60}^L(k) $ \\
    \hline
    0 &  0.4536364 & 0.4546717 & $-$0.0010352\\
    1 &  0.3676810 &  0.3674131 & 0.00026781 \\
    2   & 0.0591954 & 0.0589041  & $-$0.00029126 \\ 
   3   & 0.0017303 & 0.0018735 &  $-$0.00014319 \\
    \end{tabular}\hfill
    &
    \begin{tabular}{c|c|c|c}
$k$ &    $p_{90}^{\rm exact,L}(k) $ &  $p_{90}^{\rm approx,L}(k)$ & $\delta_{90}^L(k) $ \\
    \hline
    0 & 0.4860010  & 0.4877209 & $-$0.0017198\\
    1 &   0.2671577 &  0.2662165 &   0.0009412 \\
    2   &  0.0318105 & 0.0318451  & $-$0.0000346 \\ 
   3   &  0.0007564 & 0.0008348 &  $-$0.0000784 \\
     \end{tabular}
        \end{tabular}
      \caption{Here $\delta_N^L(k):= p_N^{\rm exact,L}(k)  -  p_N^{\rm approx,L}(k)$ }
         \label{TB1}
 \end{table}%

As for the GOE, and after fixing a value of the Laguerre parameter $a$ as well as $N$, we have access to a tabulation of
the marginal eigenvalue PDFs in the bulk. On the theory side, using a very different set of ideas than those used to prove (\ref{3.8}), now based on the tridiagonal model of \cite{DE02} and martingale arguments, it is established in \cite{Re22} that
 \begin{equation}\label{3.8L}
\lim_{N,l \to \infty}   {\pi \rho^{\rm MP}(\gamma_l) \over \log N} \Big ( x_l  -N \gamma_l \Big ) \mathop{=}\limits^{\rm d} {\tt N}[0,1],
\end{equation}
where $\gamma_l$ is such that
$
 \int_{\gamma_l}^{c_+} \rho^{\rm MP}(x) \, dx = l/N$. 
 We recognise $N \gamma_l$ as the leading order of the mean (\ref{E1}), and  $\pi^2/\log N$ as being the reciprocal of the leading order of the variance (\ref{3.5L}).
 We can therefore substitute the mean and variance into (\ref{3.8L}). With this done, and $l = {2 \over 5} N$ (then $\gamma_l
 =0.9677\dots$) we have computed the PDF for the scaled random variable, $s$ say, on the LHS of (\ref{3.8L}) for various values of $N$. Accurate agreement with the PDF of the standard normal is found in all cases. For example, with $N=60$, the absolute difference is no greater that $3 \times 10^{-3}$ in the range of $s$ between $-4$ and 4.

\subsection*{Ancillary Mathematica files}
The various tables and graphs in the text were produced by implementing the specified computational schemes as Mathematica notebooks.
The symbolic computations of the marginal distributions for the GOE reported in Appendix A use  {\tt GOEsymbolic.nb}. Numerical calculations for the probabilities $\{E_N^{\rm G}(k;(s,\infty) \}_{k=0}^N$ with $s$ fixed  as required to produce Tables \ref{TA}--\ref{T4} use {\tt GOEnumeric1.nb}, while the numerical computation of the marginal distributions in the Gaussian case, which were required to produce Figure \ref{F2} use
{\tt GOEnumeric2.nb}. The Tables \ref{T4a}--\ref{TB1} in relation to the Laguerre case use {\tt LOEnumeric.nb}. Table \ref{T7} in Appendix A relates to the LOE, and uses {\tt LOEsymbolic.nb}.

 \subsection*{Acknowledgements}
	PJF and BJS are supported
	by the Australian Research Council 
	 Discovery Project grant DP210102887, and a 2024 University of Melbourne Science Faculty small grant.
	 SK acknowledges the support provided by SERB, DST, Government of India, via Grant No.
CRG/2022/001751.
   
 \pagebreak

\section*{Appendix A}
\renewcommand{\thesection}{A} 
\setcounter{equation}{0}

In Section \ref{S2.1} a formalism to compute  the exact functional forms of the independent marginal individual eigenvalue PDFs $\{f_N^{\rm G}(k;s)\}_{k=1}^{\lfloor N/2 \rfloor}$
has been presented. Here, extending results in   \cite[Appendix 4]{Ek74}, the results of using these exact functional forms to compute the corresponding statistical quantities
 $\{\mu,\sigma,\gamma_1, \gamma_2,\gamma_3,\gamma_4\}$ (recall (\ref{2.12})) is presented in tabular form, where in keeping with \cite[Appendix 4]{Ek74} the decimals are truncated
 after the  7th place.

\begin{table}[h]
\begin{tabular}{cc|cccccc}
$N$ & $k$ & $\mu$ & $\sigma$ & $\gamma_1$ &  $\gamma_2$ & $\gamma_3$ & $\gamma_4$  \\
\hline
8 & 1 & 3.2451029 & 0.6431706 & 0.2175978 & 0.0866120 & 0.0161454 & $-$0.0388095 \\
8 & 2 & 2.1372504 & 0.5423684 & 0.0972884 & 0.0126373 & $-$0.0059766 & $-$0.00572405 \\
8 & 3 & 1.2367953 & 0.5064137 & 0.0468383& $-$0.0040771 & $-$0.0046966 & 0.0017507 \\
8 & 4 & 0.4060264 & 0.4926050  & 0.0142245 & $-$0.0089303 & $-$0.0015978 & 0.0039260 \\
{} & & & & & & & \\
9 & 1 & 3.5094346 & 0.6296768 & 0.2232643 & 0.0913421 & 0.0201166 & $-$0.0370827 \\
9 & 2 & 2.4308448 & 0.5285622 & 0.1033451 & 0.0148436 & $-$0.0054812 & $-$0.0062513 \\
9 & 3 & 1.5608761 & 0.4908939 & 0.0542856 & $-$0.0028530 & $-$0.0051481 & 0.0011737 \\
9 & 4 & 0.7660494 & 0.4741134 & 0.0241204 & $-$0.0085843 & $-$0.0026359 & 0.0036123 \\
9 & 5 &  0 & 0.4691792 & 0 & $-$0.0100241 & 0 & 0.0042217 \\
{} & & & & & & & \\
10 & 1 & 3.7575287 & 0.6179386 & 0.2279701 & 0.0953997 & 0.0236601 & $-$0.0352918 \\
10 & 2 & 2.7037563 & 0.5168106 & 0.1082262 & 0.0167597 & $-$0.0049572 & $-$0.0066215 \\
10 & 3 & 1.8588219 &  0.4780208 & 0.0600734 & $-$0.0017051 & $-$0.0054038 & 0.0006890 \\
10 & 4 & 1.0923667 & 0.4592937 & 0.2109507 & 0.0314529 & $-$0.0080085 & $-$0.0033338 \\
10 & 5 & 0.3607214 & 0.4513082 & 0.0098626 & $-$0.0102013 & $-$0.0011011& 0.0041399 \\
{} & & & & & & & \\
11 & 1 & 3.9920188 &  0.6075750 & 0.2319524 & 0.0989270 & 0.0268408 & $-$0.0335013 \\
11 & 2 & 2.9597112 & 0.5066156 & 0.11225752 & 0.0184367 & $-$0.0044339 & $-$0.0068800 \\
11 & 3 & 2.1358866 &  0.4670738 & 0.0647201 & $-$0.0006566 & $-$0.0055397 & 0.0002862 \\
11 & 4 & 1.3926567 & 0.4470071 & 0.0371329 & $-$0.0073705 & $-$0.0038169 & 0.0029094 \\
11 & 5 & 0.6881186 &   0.4369907 & 0.0171500 & $-$0.0100138 & $-$0.0018800 & 0.0039388 \\
11 & 6 & 0 & 0.4339162 & 0 & $-$0.0107352 & 0 & 0.0042171 \\
{} & & & & & & & \\ 
12 & 1 & 4.2148992 & 0.5983148 &  0.2353747 & 0.1020278 & 0.0297125 & $-$0.0317470 \\
12 & 2 & 3.2014467 & 0.4976369 & 0.1156525 & 0.0199163 & $-$0.0039259 & $-$0.0070582 \\
12 & 3 & 2.3957884 & 0.4575857 & 0.0685453 & 0.0002926 & $-$0.0056002 & $-$0.0000490 \\
12 & 4 & 1.6720878 & 0.4365640 & 0.0416807 & $-$0.0067366 & $-$0.0041593 & 0.0026001 \\
12 & 5 & 0.9897011 & 0.4251242 & 0.0227820 & $-$0.0096782 & $-$0.0024465 & 0.0037082 \\
12 & 6 & 0.3278102 & 0.4199860 & 0.0072728 & $-$0.0108093 & $-$0.0008020 & 0.0041287 
\end{tabular}
\end{table}

We take this opportunity to draw attention to some inequalities associated with the means $\mu = \mu_{N,k}$. In relation to this, we require the fact that upon appropriate Householder similarity transformations, GOE matrices can be demonstrated to be similar to certain symmetric random tridiagonal matrices $T_N$ \cite{Tr84}. The $T_N$ have independent standard Gaussian entries on the diagonal, and leading off diagonal entries given by the random variables $\{ \tilde{\chi}_{N-j} \}_{j=1,\dots,N-j} $, where $\tilde{\chi}_k$
 is the square root of the usual $\chi_k^2$ random variable, scaled by $1/\sqrt{2}$. In particular $\mathbb E \, T_N$  is the symmetric tridiagonal matrix with zero on its diagonal entries and leading off diagonal entries equal to $\{ {1 \over \sqrt{2}} \sqrt{N-j} \}_{j=1,\dots,N-1}$. From the implied three term recurrence one can check that $\det (x \mathbb I_N - \mathbb E \, T_N) \propto H_N(x)$, where $H_N(x)$ denotes the Hermite polynomial of degree $N$, telling us that eigenvalues of $\mathbb E \, T_N$ are equal to the zeros of this Hermite polynomial.
 
 On the other hand, for general random real symmetric matrices $T_N$ a theorem of Cacoullos and Olkin \cite[Corollary 2.8]{CO65} gives that
 \begin{equation}\label{C6}
\mathbb E \,\lambda_1(T_N) + \cdots + \mathbb E \, \lambda_k(T_N) \ge  
\lambda_1(\mathbb E \, \mathbb T_N) + \cdots + \lambda_k(\mathbb E \, T_N).
\end{equation}
By computing the zeros of the Hermite polynomials, and comparing with the $\mu_{N,k}$ in the table, we can readily verify (\ref{C6}). In fact in doing so, one observes the stronger interlacing inequality
 \begin{equation}\label{C7}
 \mathbb E \,\lambda_k(T_N)  >  
\lambda_k(\mathbb E \, \mathbb T_N)>\mathbb E \,\lambda_{k-1}(T_N), \qquad (k=1,\dots,\lfloor N/2 \rfloor)
\end{equation}
exhibited by the listed $\mu_{N,k}$. 

An analogous property is observed in the Laguerre case. The tridiagonal matrix $T_N$ similar to a LOE matrix has the property that $\det (x \mathbb I_N - \mathbb E \, \mathbb T_N) \propto L_N^{2a-1}(x)$, where $L_N^{2a-1}(x)$ denotes the Laguerre polynomial of degree $N$ with parameter $2a-1$ \cite[Table 2(a) with $\gamma \mapsto 2a$]{DE05}. The following table presents the means of the marginal eigenvalue PDFs $\{f_N^{\rm L}(k;s)\}_{k=1}^{N}$ obtained by numerical integration of its exact functional form, which is computed according to the formalism in Section 2.2. By comparing the means with the zeros of $L_N^{2a-1}(x)$, one can verify the validity of the interlacing \eqref{C7} for each $k=1,\dots,N$.

\begin{table}[h]
\centering
\begin{tabular}{cc|ccccc}
 & $N$ & $2$ & $3$ & $4$ & $5$ & $6$\\
$a$ & $k$ &  &  &  &    &   \\
\hline
4 & 1 & 15.0634920 & 19.4986342 & 23.7003816 & 27.7874859 & 31.8100369  \\
4 & 2 & 6.9365079 & 10.9999999 & 14.7379489 & 18.3726752 & 21.9688258  \\
4 & 3 &           & 5.5013657 & 8.9489690& 12.2197905 & 15.4601777 \\
4 & 4 &           &           & 4.6127004 & 7.6273247 & 10.5502764  \\
4 & 5 &           &           &           & 3.9927234 & 6.6808343  \\
4 & 6 &           &           &           &           & 3.5298487 \\
{} & & & & & & \\
7/2 & 1 & 13.8656315 & 18.1848768 & 22.3053570 & 26.3308207 & 30.3044360  \\
7/2 & 2 & 6.1343684 & 10.0000000 & 13.6136374 & 17.1577682 & 20.6833885  \\
7/2 & 3 &           & 4.8151231 & 8.0729350& 11.2189206 & 14.3657513 \\
7/2 & 4 &           &           & 4.0080704 & 6.8422317 & 9.6424231  \\
7/2 & 5 &           &           &           & 3.4502585 & 5.9670831  \\
7/2 & 6 &           &           &           &           & 3.0369177 \\
\end{tabular}
\caption{Means of the LOE marginal density $f_N^L(k;s)$}
\label{T7}
\end{table}

%
%
%
%
%
%
%

\pagebreak

  \providecommand{\bysame}{\leavevmode\hbox to3em{\hrulefill}\thinspace}
\providecommand{\MR}{\relax\ifhmode\unskip\space\fi MR }
\providecommand{\MRhref}[2]{%
  \href{http://www.ams.org/mathscinet-getitem?mr=#1}{#2}
}
\providecommand{\href}[2]{#2}


\begin{thebibliography}{10}

\bibitem{AE06}
A. Aazami and R. Easther, \emph{Cosmology from random multifield potentials}, Journal of Cosmology
and Astroparticle Physics \textbf{2006} (2006): 013.


\bibitem{ABC13}
A. Aunger, G. Ben Arous and J. \v{C}ern\'{y}, \emph{Random matrices and complexity of spin glasses},
Comm. Pure Appl. Math. \textbf{66} (2013), 165--201.

\bibitem{AD22}
J.M. Aza\"{i}s and C.~Delmas, \emph{Mean number and correlation function of critical points of
isotropic Gaussian fields and some results on GOE random matrices}, Stochastic Processes and their
Applications, \textbf{150} (2022), 411-445.

\bibitem{BDJ98}
J.~Baik, P.~Deift, and K.~Johansson, \emph{On the distribution of the length of
  the longest increasing subsequence of random permutations}, J. Amer. Math.
  Soc. \textbf{12} (1999), 1119--1178.
  
  \bibitem{BF03}
A.~Borodin and P.J. Forrester, \emph{Increasing subsequences and the
  hard-to-soft transition in matrix ensembles}, J.Phys. A \textbf{36} (2003),
  2963--2981.


\bibitem{Bo08}
  F.~Bornemann, \emph{On the numerical evaluation of {F}redholm determinants}, Math.
  Comp. \textbf{79} (2010), 871--915.

 \bibitem{Bo10}
  F.~Bornemann,  On the numerical evaluation of distributions in random
  matrix theory: a review, {\it Markov Processes Relat. Fields} 16  (2010),  803--866.
  
  \bibitem{Bo23}  
 F.~ Bornemann,  \emph{A Stirling-type formula for the distribution of the length of longest increasing subsequences}, Found Comput. Math. 
 \textbf{24} (2024), 915--953.
 
 \bibitem{Bo24}
 F.~Bornemann, \emph{Asymptotic expansions relating to the distribution of the length of longest increasing subsequences}, Forum Math. Sigma. 12:e36. (2024) doi:10.1017/fms.2024.13 
 
 \bibitem{Bo24a}
 F.~Bornemann, \emph{Asymptotic expansions of the limit laws of Gaussian and Laguerre (Wishart) ensembles at the soft edge}, arXiv:2403.07628
 
   \bibitem{BFM17}
F.~Bornemann, P.J. Forrester, and A.~Mays, \emph{Finite size effects for
  spacing distributions in random matrix theory: circular ensembles and Riemann
  zeros}, Stud. Appl. Math. \textbf{138} (2017), 401--437.

\bibitem{BEMN11}
G. Borot, B. Eynard, S. N. Majumdar, and C. Nadal, \emph{Large deviations
of the maximal eigenvalue of random matrices}. J. Stat. Mech. Theory
Exp.,  (2011), P11024.

\bibitem{dB55}
N.G. de~Bruijn, \emph{On some multiple integrals involving determinants}, J.
  Indian Math. Soc. \textbf{19} (1955), 133--151.
  
 \bibitem{BF23}
        S.-S.~Byun and P.J.~Forrester,   \emph{Progress on the study of the Ginibre ensembles},
  KIAS Springer Series in Mathematics \textbf{3}, Springer, 2024.

  
  \bibitem{BL24}
  S.-S. Byun and Y.-W. Lee, \emph{Finite size corrections for real eigenvalues of the elliptic Ginibre matrices},
  Random Matrices: Theory and
Applications \textbf{13} (2024) 2450005.

\bibitem{CO65}
T.~Cacoullos and I.~Olkin,
\emph{On the bias of functions of characteristic roots of a random matrix}, Biometrika \textbf{52} (1965), 87--94.
  
\bibitem{CGG00}  
  A. Cavagna, J. P. Garrahan and I. Giardina, \emph{Index distribution of random matrices with an application to disordered systems}, Phys. Rev. B \textbf{61}, 3960 (2000)

\bibitem{Ch14}
M. Chiani, \emph{Distribution of the largest eigenvalue for real Wishart and Gaussian random
matrices and a simple approximation for the Tracy-Widom distribution}, J. Multivariate
Anal. \textbf{129} (2014) 69--81.

\bibitem{CL95}
O.~Costin and J.L. Lebowitz, \emph{Gaussian fluctuations in random matrices},
  Phys. Rev. Lett. \textbf{75} (1995), 69--72.

   \bibitem{Da72a}
A.W. Davis, \emph{On the marginal distributions of the latent roots of the
  multivariable beta matrix}, Ann. Math. Statist. \textbf{43} (1972),
  1664--1669.

\bibitem{Da72b}
A.W. Davis, \emph{
  On the distributions of the latent roots and traces of certain random matrices}, J. Multivariate
  Anal. \textbf{2} (1972), 189--200.
  
\bibitem{DS17}  
  A.~Dea\~{n}o and N.~Simm, \emph{On the probability of positive-definiteness in the
gGUE via semi-classical Laguerre polynomials}, J. Approx. Theory, \textbf{219} (2017), 44--59.
  
  \bibitem{DE02}
I.~Dumitriu and A.~Edelman, \emph{Matrix models for beta ensembles}, J. Math.
  Phys. \textbf{43} (2002), 5830--5847.


\bibitem{DE05}
  I. Dumitriu and A. Edelman, \emph{Eigenvalues of Hermite and Laguerre ensembles: large beta
asymptotics}, Ann. Inst. H. Poincar\'e Probab. Statist., \textbf{41} (2005), 1083--1099.

  
  
    \bibitem{Ek74}
  S.R.~Eckert, \emph{Distributions of the individual ordered roots of random matrices}, PhD. thesis,
  University of Adelaide, 1974.

  \bibitem{FTW23}
R.~Feng, G.~Tian and D.~Wei,
\emph{The Berry-Esseen theorem for circular $\beta$-ensemble}, 
  Annals Appl. Probab. \textbf{33} (2023), 5050--5070.
  
  \bibitem{Fo93a}
P.J. Forrester, \emph{The spectrum edge of random matrix ensembles}, Nucl. Phys. B
  \textbf{402} (1993), 709--728.


\bibitem{Fo10}
P.J. Forrester,  \emph{Log-gases and random matrices}, Princeton University Press,
  Princeton, NJ, 2010.
  
      \bibitem{Fo23}
P.J. Forrester, \emph{A review of exact results for fluctuation formulas in random matrix theory},
Probab. Surveys \textbf{20} (2023), 170--225. 

\bibitem{Fo24}
P.J. Forrester, \emph{Local central limit theorem for real eigenvalue fluctuations of elliptic GinOE matrices},
Electron. Commun. Probab. \textbf{29} (2024), 1--11.


  
  \bibitem{FK19}
P.J. Forrester and S.~Kumar, 
\emph{Recursion scheme for the largest $\beta$-Wishart-Laguerre
eigenvalue and Landauer conductance in quantum transport}, J. Phys. A \textbf{52} (2019), 42LT02. 

  \bibitem{FK24}
  P.J. Forrester and S.~Kumar, 
\emph{Computation of marginal eigenvalue distributions in the Laguerre and Jacobi $\beta$ ensembles},
arXiv:2402.16069.

  \bibitem{FL14} 
  P. Forrester, J. Lebowitz,  \emph{Local central limit theorem for determinantal point
processes}, J. Stat. Phys. \textbf{157}  (2014), 60--69.

   \bibitem{FLT20}
  P.J. Forrester, S.-H.~Li and A.K.~Trinh,  \emph{Asymptotic correlations with corrections for the circular Jacobi
  $\beta$-ensemble}, J. Approximation Th.   \textbf{271} (2021), 105633.

  \bibitem{FM23}
  P.J. Forrester and A.~Mays, \emph{Finite size corrections relating to distributions of the
  length of longest increasing subsequences}, Adv. Applied Math. \textbf{145} (2023), 102482.
  
    \bibitem{FR01}
P.J. Forrester and E.M. Rains, \emph{Inter-relationships between orthogonal, unitary and symplectic matrix
ensembles}, In P.M. Bleher and A.R. Its, editors,
\emph{Random matrix models and their applications}. volume 40 of \emph{Mathematical Sciences 
Research Institute Publications}, pages 171-208. Cambridge University Press, United Kingdom, 2001.
  
  \bibitem{FT18}
P.J.~Forrester and A.K.~Trinh. 
\newblock \emph{Functional form for the leading correction to the distribution of the largest eigenvalue in the GUE and LUE}. \newblock J. Math. Phys., 59(5) (2018), 053302.
  
  \bibitem{FT19a} P. J. Forrester and A. K.~Trinh,
\emph{Optimal soft edge scaling variables for the Gaussian and Laguerre even ensembles}, Nucl. Phys. B 938 (2019), 621--639.



\bibitem{Gu05}
J.~Gustavsson, \emph{Gaussian fluctuations in the {GUE}}, Ann. l'Inst. Henri
  Poincar\'e (B) \textbf{41} (2005), 151--178.


  
  
  \bibitem{Ja75} A.T. James, \emph{Special functions of matrix and single argument in statistics}, in Theory and Applications of Special Functions (R. A. Askey, Ed.), Academic, New York, 1975, pp. 497--520.
  
  \bibitem{Ki08}
R.~Killip, \emph{Gaussian fluctuations for $\beta$ ensembles}, Int. Math. Res.
  Not. \textbf{2008} (2008), rnn007.
  
  \bibitem{Ma11}
A.~Mays, \emph{A geometrical triumvirate of real random matrices}, Ph.D.
  thesis, University of Melbourne, 2012.
  
  
  \bibitem{MNSV11} 
   S.N. Majumdar, C. Nadal, A. Scardicchio, and P. Vivo, \emph{How many eigenvalues of a gaussian random matrix are positive?}
    Phys. Rev. E, \textbf{83} (2011), 04110.

\bibitem{MV12}
    S.N. Majumdar and P. Vivo, \emph{Number of relevant directions in principal component analysis and
Wishart random matrices}, Phys. Rev. Lett., \textbf{108} (2012), 200601.

\bibitem{MP24}
A.M.~Mathai and S.B.~Provost, \emph{The exact density of the eigenvalues of the Wishart and matrix-variate gamma and beta random variables}, Mathematics.  \textbf{12} (2024), 2427. \\
https://doi.org/10.3390/math12152427 
  
  \bibitem{Me67}
M.L. Mehta, \emph{Random matrices and the statistical theory of energy levels},
  Academic Press, New York, 1967.
  
  \bibitem{MH06}
  L. Mersini-Houghton, \emph{Can we predict Lambda for the non-SUSY sector of the landscape?} Class. Quant. Grav.
  \textbf{22} (2005),  3481.
  
   \bibitem{OR10} 
  S. O'Rourke, \emph{Gaussian fluctuations of eigenvalues in Wigner random matrices}, J. Stat.
Phys., \textbf{138} (2010), 1045--1066.

    \bibitem{PS11}
L.~Pastur and M.~Shcherbina, \emph{Eigenvalue distribution of large random
  matrices}, American Mathematical Society, Providence, RI, 2011.

\bibitem{PS16}
		A.~Perret and G.~Schehr, \emph{Finite N corrections to the limiting distribution of the smallest eigenvalue of Wishart complex matrices},
		 Random Matrices: Theory and Applications, \textbf{5} (2016), 1650001.

 \bibitem{Re22}
 T.~Reeves, \emph{Central limit theorem for fluctuation of eigenvalues of real Wishart matrices}, PhD thesis, Cornell University, 2022.


 \bibitem{Sh15}
M.~Shcherbina, \emph{Fluctuations of the eigenvalue number in the fixed interval for $\beta$-models with $\beta = 1,2,4$},
pp.~131--146 in ``Theory and Applications in Mathematical Physics", Ed.~E.~Agliari et al., World Scientific, 2015.

 \bibitem{SLMS21}
N.R.~Smith, P. Le Doussal, S.N.~Majumdar, and G.~Schehr, \emph{Counting statistics for non-interacting
fermions in a d-dimensional potential}, Phys. Rev. E \textbf{103} (2021), L030105.


  \bibitem{So00}
A. Soshnikov, \emph{Determinantal random point fields},
Russian Math. Surveys, \textbf{55} (2000),   923--975.

\bibitem{SE16} Stack Exchange, \emph{Compute numerical Pfaffians of matrices efficiently?}, \\
https://mathematica.stackexchange.com/questions/125794 (2016)

\bibitem{St73}
G.W. Stewart, \emph{Introduction to matrix computations}, Academic Press, New
York, 1973.
  
  
  \bibitem{TW94a}
C.A. Tracy and H.~Widom, \emph{Level-spacing distributions and the {Airy} kernel}, Commun.
  Math. Phys. \textbf{159} (1994), 151--174.

    \bibitem{Tr84}
H.F.~Trotter, \emph{Eigenvalue distributions of large {Hermitian} matrices:
  {Wigner's} semi-circle law and a theorem of {Kac}, {Murdock} and {Szeg\"o}},
  Adv. Math. \textbf{54} (1984), 67--82.
  
   \bibitem{Wi12}
  M. Wimmer, \emph{Algorithm 923: Efficient numerical computation of the Pfaffian for
dense and banded skew-symmetric matrices}, ACM Trans. Math. Softw. \textbf{38}  (2012), 30.

 \bibitem{WF12}
N.S.~Witte and P.J.~Forrester,  \emph{On the variance of the index for the Gaussian unitary ensemble} Random Matrices: Theory and
Applications \textbf{1} (2012) 1250010.

  
  
  \end{thebibliography}
  \end{document}